\documentclass[12pt]{article}

\usepackage{graphicx}
\usepackage{float}
\usepackage{cite}
\usepackage{amsfonts}
\usepackage{amssymb}
\usepackage{xcolor}

\def\gtwid{\mathrel{\raise.3ex\hbox{$>$\kern-.75em\lower1ex\hbox{$\sim$}}}}
\def\ltwid{\mathrel{\raise.3ex\hbox{$<$\kern-.75em\lower1ex\hbox{$\sim$}}}}
\def\square{\kern1pt\vbox{\hrule height 1.2pt\hbox{\vrule width 1.2pt\hskip 3pt
   \vbox{\vskip 6pt}\hskip 3pt\vrule width 0.6pt}\hrule height 0.6pt}\kern1pt}

\begin{document}

\begin{titlepage}

\begin{flushright}
UFIFT-QG-22-01 , CCTP-2022-3
\end{flushright}

\vskip 1cm

\begin{center}
{\bf How Inflationary Gravitons Affect the Force of Gravity}
\end{center}

\vskip .5cm

\begin{center}
L. Tan$^{1\star}$, N. C. Tsamis$^{2\dagger}$ and R. P. Woodard$^{1\ddagger}$
\end{center}

\vskip .5cm

\begin{center}
\it{$^{1}$ Department of Physics, University of Florida,\\
Gainesville, FL 32611, UNITED STATES}
\end{center}

\begin{center}
\it{$^{2}$ Institute of Theoretical Physics \& Computational Physics, \\
Department of Physics, University of Crete, \\
GR-710 03 Heraklion, HELLAS}
\end{center}

\vspace{0.5cm}

\begin{center}
ABSTRACT
\end{center}
We employ an unregulated computation the graviton self-energy from gravitons 
on de Sitter background to infer the renormalized result. This is used 
to quantum-correct the linearized Einstein equation. We solve this equation
for the potentials which represent the gravitational response to static, 
point mass. We find large spatial and temporal logarithmic corrections to 
the Newtonian potential and to the gravitational shift. Although suppressed
by a minuscule loop-counting parameter, these corrections cause perturbation
theory to break down at large distances and late times. Another interesting
fact is that gravitons induce up to three large logarithms whereas a loop of
massless, minimally coupled scalars produces only a single large logarithm.
This is in line with corrections to the graviton mode function: a loop of
gravitons induces two large logarithms whereas a scalar loop gives none.

\begin{flushleft}
PACS numbers: 04.50.Kd, 95.35.+d, 98.62.-g
\end{flushleft}

\vskip .5cm

\begin{flushleft}
$^{\star}$ e-mail: ltan@ufl.edu \\
$^{\dagger}$ e-mail: tsamis@physics.uoc.gr \\
$^{\ddagger}$ e-mail: woodard@phys.ufl.edu
\end{flushleft}

\end{titlepage}

\section{Introduction}

A key prediction of primordial inflation is that virtual gravitons of
cosmological scale are ripped out of the vacuum \cite{Starobinsky:1979ty,
Starobinsky:1985ww}. The occupation number for each wave vector $\vec{k}$
is staggering,
\begin{equation}
N(\eta,k) = \frac{\pi \Delta^2_{h}(k)}{64 G k^2} \times a^2(\eta) \; ,
\end{equation}
where $\Delta^2_{h}(k)$ is the tensor power spectrum, $G$ is Newton's
constant and $a(\eta)$ is the scale factor at conformal time $\eta$. Our
goal is to study how these gravitons change the force of gravity.

We can describe the background geometry of cosmology in conformal 
coordinates,
\begin{equation}
ds^2 = a^2(\eta) \Bigl[-d\eta^2 + d\vec{x} \!\cdot\! d\vec{x}\Bigr] \qquad
\Longrightarrow \qquad H \equiv \frac{a'}{a^2} \quad , \quad \epsilon \equiv 
-\frac{H'}{a H^2} \; , \label{geometry}
\end{equation}
where $H(\eta)$ is the Hubble parameter and $\epsilon(\eta)$ is the first
slow roll parameter. A reasonable paradigm for inflation is provided by the 
special case of de Sitter ($\epsilon = 0$, constant $H$ and $a(\eta) = 
-1/H\eta$), which is tempting because there are analytic expressions for
the graviton propagator \cite{Tsamis:1992xa,Woodard:2004ut} and because 
there is no mixing between gravitons and the matter fields that drive 
inflation \cite{Iliopoulos:1998wq,Abramo:2001dc}. One quantum-corrects the
linearized Einstein equation using the graviton self-energy $-i 
[\mbox{}^{\mu\nu} \Sigma^{\rho\sigma}](x;x')$ which is the 1PI (one particle
irreducible) 2-graviton function,
\begin{equation}
\mathcal{D}^{\mu\nu\rho\sigma} h_{\rho\sigma}(x) - \int \!\!
d^4x' \Bigl[\mbox{}^{\mu\nu} \Sigma^{\rho\sigma}\Bigr](x;x') h_{\rho\sigma}(x')
= \frac12 \kappa T^{\mu\nu}_{\rm lin}(x) \; . \label{Einsteineqn}
\end{equation}
Here $\kappa^2 \equiv 16 \pi G$ is the loop-counting parameter, $h_{\mu\nu} 
\equiv (g_{\mu\nu} - a^2 \eta_{\mu\nu})/\kappa$ is the graviton field, 
$T^{\mu\nu}_{\rm lin}(x)$ is the linearized stress tensor and $\mathcal{D}^{
\mu\nu\rho\sigma}$ is the graviton kinetic operator in the same gauge that 
was used to compute $-i [\mbox{}^{\mu\nu} \Sigma^{\rho\sigma}](x;x')$. Our 
two aims in this work are (1) to infer a fully renormalized result for 
$-i[\mbox{}^{\mu\nu} \Sigma^{\rho\sigma}](x;x')$ at one loop from an old 
computation \cite{Tsamis:1996qk} that was made without regularization, and
(2) to work out one loop corrections to the gravitational response to a point 
mass. 

There are four sections to this paper, of which this Introduction is the 
first. Section 2 describes our procedure for extracting the renormalized 
self-energy from the unregulated result, with technical details consigned to 
an Appendix. Section 3 solves (\ref{Einsteineqn}) for one loop corrections 
to the gravitational potentials induced by a point mass. Our conclusions 
comprise section 4.

\section{Quantum Linearized Einstein Equation}

This section derives an explicit expression for the quantum-corrected
Einstein equation (\ref{Einsteineqn}). Our first tasks are specifying the
gauge-fixed kinetic operator $\mathcal{D}^{\mu\nu\rho\sigma}$, explaining
how we represent the tensor structure of the graviton self-energy, and 
giving $3+1$ decompositions of both. The main part of this section is 
describing the process through which we infer most of the renormalized, 
Schwinger-Keldysh result for the graviton self-energy from an unregulated, 
noncoincident computation \cite{Tsamis:1996qk}. At the section's end we
give a direct, dimensionally regulated computation of the local 4-point 
contribution, and we discuss the need for a fully dimensionally regulated 
calculation.

\subsection{$3+1$ Decomposition}

In the simplest gauge and $D = 3+1$ dimensions, the gauge-fixed kinetic 
operator takes the form \cite{Tsamis:1992xa,Woodard:2004ut},
\begin{equation}
\mathcal{D}^{\mu\nu\rho\sigma} = \frac12 \eta^{\mu (\rho} \eta^{\sigma )\nu}
D_A - \frac14 \eta^{\mu\nu} \eta^{\rho\sigma} D_A + 2 a^4 H^2 \delta^{(\mu}_{~~0}
\eta^{\nu ) (\rho} \delta^{\sigma)}_{~~0} \; . \label{kinop}
\end{equation}
Here $D_A$ is the massless, minimally coupled scalar kinetic operator,
\begin{equation}
D_A = -a^2 \Bigl[ \partial_0^2 + 2 a H \partial_0 - \nabla^2\Bigr] = 
\partial^{\mu} a^2 \partial_{\mu} \; . \label{DAdef}
\end{equation}
The $3+1$ decomposition of $\mathcal{D}^{\mu\nu\rho\sigma} h_{\rho\sigma}$ is,
\begin{eqnarray}
\mathcal{D}^{00\rho\sigma} h_{\rho\sigma} & = & \frac14 D_A (h_{00} + h_{kk})
- 2 a^4 H^2 h_{00} \; , \qquad \label{D00} \\
\mathcal{D}^{0i\rho\sigma} h_{\rho\sigma} & = & -\frac12 D_B h_{0i} \; , 
\label{D0i} \\
\mathcal{D}^{ij\rho\sigma} h_{\rho\sigma} & = & \frac12 D_A \Bigl[ h_{ij} +
\frac12 \delta_{ij} (h_{00} - h_{kk})\Bigr] \; , \qquad \label{Dij}
\end{eqnarray}
where $D_B$ stands for the kinetic operator of a massless, conformally coupled 
scalar,
\begin{equation}
D_B = -a^2 \Bigl[ \partial_0^2 + 2 a H \partial_0 - \nabla^2 + 2 a^2 H^2\Bigr] 
= a \partial^2 a \; . \label{DBdef}
\end{equation}
Note that adding (\ref{D00}) and the trace of (\ref{Dij}) gives a relation for 
$h_{00}$,
\begin{equation}
\Bigl( \mathcal{D}^{00\rho\sigma} + \mathcal{D}^{kk\rho\sigma}\Bigr) 
h_{\rho\sigma} = D_B h_{00} \; . \label{h00eqn}
\end{equation}

Using general tensor analysis on a general cosmological background 
(\ref{geometry}), we can represent the graviton self-energy as a sum of 21 
tensor differential operators $[\mbox{}^{\mu\nu} \mathcal{D}^{\rho\sigma}]$ 
acting on scalar functions of $\eta$, $\eta'$ and $\Vert \vec{x} - \vec{x}'
\Vert$ \cite{Tan:2021ibs},
\begin{equation}
-i\Bigl[\mbox{}^{\mu\nu} \Sigma^{\rho\sigma}\Bigr](x;x') = \sum_{i=1}^{21}
\Bigl[\mbox{}^{\mu\nu} \mathcal{D}_{i}^{\rho\sigma}\Bigr] \!\times\! T^i(x;x') 
\; . \label{initialrep}
\end{equation}
The 21 basis tensors are constructed from $\delta^{\mu}_{~0}$, the spatial 
part of the Min\-kow\-ski metric $\overline{\eta}^{\mu\nu} \equiv \eta^{\mu\nu} 
+ \delta^{\mu}_{~0} \delta^{\nu}_{~0}$ and the spatial derivative operator 
$\overline{\partial}^{\mu} \equiv \partial^{\mu} + \delta^{\mu}_{~0} \partial_0$. 
These 21 tensors are listed in Table~\ref{Tbasis}.
\begin{table}[H]
\setlength{\tabcolsep}{8pt}
\def\arraystretch{1.5}
\centering
\begin{tabular}{|@{\hskip 1mm }c@{\hskip 1mm }||c||c|c||c|c|}
\hline
$i$ & $[\mbox{}^{\mu\nu} \mathcal{D}^{\rho\sigma}_i]$ & $i$ & $[\mbox{}^{\mu\nu} 
\mathcal{D}^{\rho\sigma}_i]$ & $i$ & $[\mbox{}^{\mu\nu} \mathcal{D}^{\rho\sigma}_i]$ \\
\hline\hline
1 & $\overline{\eta}^{\mu\nu} \overline{\eta}^{\rho\sigma}$ & 8 & $\overline{\partial}^{\mu} 
\overline{\partial}^{\nu} \overline{\eta}^{\rho\sigma}$ & 15 & $\delta^{(\mu}_{~~0} 
\overline{\partial}^{\nu)} \delta^{\rho}_{~0} \delta^{\sigma}_{~0}$ \\
\hline
2 & $\overline{\eta}^{\mu (\rho} \overline{\eta}^{\sigma) \nu}$ & 9 & $\delta^{(\mu}_{~~0} 
\overline{\eta}^{\nu) (\rho} \delta^{\sigma)}_{~~0}$ & 16 & $\delta^{\mu}_{~0} \delta^{\nu}_{~0} 
\overline{\partial}^{\rho} \overline{\partial}^{\sigma}$ \\
\hline
3 & $\overline{\eta}^{\mu\nu} \delta^{\rho}_{~0} \delta^{\sigma}_{~0}$ & 10 & 
$\delta^{(\mu}_{~~0} \overline{\eta}^{\nu) (\rho} \overline{\partial}^{\sigma)}$ & 17 & 
$\overline{\partial}^{\mu} \overline{\partial}^{\nu} \delta^{\rho}_{~0} \delta^{\sigma}_{~0}$ \\
\hline
4 & $\delta^{\mu}_{~0} \delta^{\nu}_{~0} \overline{\eta}^{\rho\sigma}$ & 11 & 
$\overline{\partial}^{(\mu} \overline{\eta}^{\nu) (\rho} \delta^{\sigma)}_{~~0}$ & 18 
& $\delta^{(\mu}_{~~0} \overline{\partial}^{\nu)} \delta^{(\rho}_{~~0} 
\overline{\partial}^{\sigma)}$ \\
\hline
5 & $\overline{\eta}^{\mu\nu} \delta^{(\rho}_{~~0} \overline{\partial}^{\sigma)}$ & 12 & 
$\overline{\partial}^{(\mu} \overline{\eta}^{\nu)(\rho} \overline{\partial}^{\sigma)}$ & 
19 & $\delta^{(\mu}_{~~0} \overline{\partial}^{\nu)} \overline{\partial}^{\rho} 
\overline{\partial}^{\sigma}$ \\
\hline
6 & $\delta^{(\mu}_{~~0} \overline{\partial}^{\nu)} \overline{\eta}^{\rho\sigma}$ & 13 & 
$\delta^{\mu}_{~0} \delta^{\nu}_{~0} \delta^{\rho}_{~0} \delta^{\sigma}_{~0}$ & 20 & 
$\overline{\partial}^{\mu} \overline{\partial}^{\nu} \delta^{(\rho}_{~~0} 
\overline{\partial}^{\sigma)}$ \\
\hline
7 & $\overline{\eta}^{\mu\nu} \overline{\partial}^{\rho} \overline{\partial}^{\sigma}$ & 
14 & $\delta^{\mu}_{~0} \delta^{\nu}_{~0} \delta^{(\rho}_{~~0} \overline{\partial}^{\sigma)}$ 
& 21 & $\overline{\partial}^{\mu} \overline{\partial}^{\nu} \overline{\partial}^{\rho} 
\overline{\partial}^{\sigma}$ \\
\hline
\end{tabular}
\caption{\footnotesize The 21 basis tensors used in expression (\ref{initialrep}). 
The pairs $(3,4)$, $(5,6)$, $(7,8)$, $(10,11)$, $(14,15)$, $(16,17)$ and $(19,20)$ 
are related by reflection.}
\label{Tbasis}
\end{table}
\noindent Table~\ref{ReflectionT} gives the 7 pairs of the $T^i(x;x')$ which are related 
by reflection invariance, $-i [\mbox{}^{\mu\nu} \Sigma^{\rho\sigma}](x;x') = -i 
[\mbox{}^{\rho\sigma} \Sigma^{\mu\nu}](x';x)$.
\begin{table}[H]
\setlength{\tabcolsep}{8pt}
\def\arraystretch{1.5}
\centering
\begin{tabular}{|@{\hskip 1mm }c@{\hskip 1mm }||c||c|c|}
\hline
$i$ & Relation & $i$ & Relation \\
\hline\hline
$3,4$ & $T^4(x;x') = +T^3(x';x)$ & $14,15$ & $T^{15}(x;x') = -T^{14}(x';x)$ \\
\hline
$5,6$ & $T^6(x;x') = -T^5(x';x)$ & $16,17$ & $T^{17}(x;x') = +T^{16}(x';x)$ \\
\hline
$7,8$ & $T^8(x;x') = +T^7(x';x)$ & $19,20$ & $T^{20}(x;x') = -T^{19}(x';x)$ \\
\hline
$10,11$ & 
$T^{11}(x;x') = -T^{10}(x';x)$ & $$ & $$ \\
\hline
\end{tabular}
\caption{\footnotesize Scalar coefficient functions in expression (\ref{initialrep})
which are related by reflection.}
\label{ReflectionT}
\end{table}

The $3+1$ decomposition of $[\mbox{}^{\mu\nu} \Sigma^{\rho\sigma}](x;x') 
h_{\rho\sigma}(x')$ is,
\begin{eqnarray}
\lefteqn{ \Bigl[\mbox{}^{00} \Sigma^{\rho\sigma}\Bigr] h_{\rho\sigma} \longrightarrow
i T^4 h_{kk} + i T^{13} h_{00} + i T^{14} h_{0k , k} + i T^{16} h_{k \ell , k\ell} 
\; , } \label{Sigma00} \\
\lefteqn{ \Bigl[\mbox{}^{0i} \Sigma^{\rho\sigma}\Bigr] h_{\rho\sigma} \longrightarrow
\frac{i}{2} \partial_i \Bigl[T^6 h_{kk} + T^{15} h_{00} + T^{18} h_{0k ,k} + T^{19}
h_{k\ell ,k\ell}\Bigr] } \nonumber \\
& & \hspace{8cm} + \frac{i}{2} T^9 h_{0i} + \frac{i}{2} T^{10} h_{i k , k} \; , 
\qquad \label{Sigma0i} \\
\lefteqn{ \Bigl[\mbox{}^{ij} \Sigma^{\rho\sigma}\Bigr] h_{\rho\sigma} \longrightarrow
i \delta_{ij} \Bigl[ T^1 h_{kk} + T^3 h_{00} + T^5 h_{0k ,k} + T^7 h_{k\ell ,k \ell}
\Bigr] + i T^2 h_{ij} } \nonumber \\
& & \hspace{-0.5cm} + i \partial_{( i} \Bigl[ T^{11} h_{j) 0} \!+\! T^{12}
h_{j ) k, k}\Bigr] + i \partial_i \partial_j \Bigl[ T^8 h_{kk} \!+\! T^{17} h_{00}
\!+\! T^{20} h_{0k ,k} \!+\! T^{21} h_{k\ell ,k\ell}\Bigr] . \qquad \label{Sigmaij}
\end{eqnarray}
Some of these relations were simplified using transition invariance to partially 
integrate spatial derivatives from the coefficient functions $T^i(x;x')$ onto the 
graviton field.

\subsection{The Quantum Correction}

\begin{figure}[H]
\centering
\includegraphics[width=11cm]{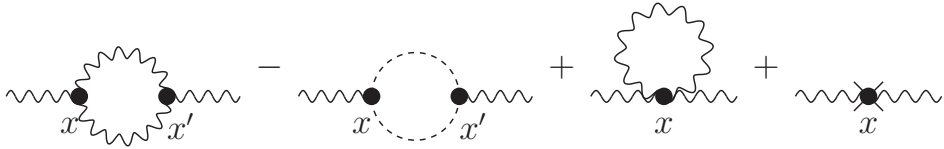}
\caption{\footnotesize Diagrams contributing to the one loop graviton 
self-energy, shown in the same order, left to right, as the four contributions 
to (\ref{operatorexpr}). Graviton lines are wavy and ghost lines are dashed.}
\label{diagrams}
\end{figure}

Suppose that $S[g]$ stands for the classical action, with ghost and gauge fixing
action $S_g[h,\overline{\theta},\theta]$, and counterterms $\Delta S[g]$. We can 
give an analytic expression for the one loop graviton self-energy using an 
expectation value of variations of these actions,
\begin{eqnarray}
\lefteqn{ -i \Bigl[\mbox{}^{\mu\nu} \Sigma^{\rho\sigma}\Bigr](x;x') = 
\Biggl\langle \Omega \Biggl\vert T^*\Biggl[ \Bigl[\frac{i \delta S[g]}{\delta 
h_{\mu\nu}(x)} \Bigr]_{h h} \Bigl[ \frac{i \delta S[g]}{\delta h_{\rho\sigma}(x')}
\Bigr]_{h h} + \Bigl[\frac{i \delta S[g]}{\delta h_{\mu\nu}(x)} \Bigr]_{
\overline{\theta} \theta} } \nonumber \\
& & \hspace{0.3cm} \times \Bigl[ \frac{i \delta S[g]}{\delta h_{\rho\sigma}(x')}
\Bigr]_{\overline{\theta} \theta} + \Bigl[ \frac{i \delta^2 S[g]}{\delta 
h_{\mu\nu}(x) \delta h_{\rho\sigma}(x')} \Bigr]_{hh} + \Bigl[ \frac{i \delta^2 
\Delta S[g]}{\delta h_{\mu\nu}(x) \delta h_{\rho\sigma}(x')} \Bigr]_{1} \Biggr] 
\Biggr\vert \Omega \Biggr\rangle . \qquad \label{operatorexpr}
\end{eqnarray}
The $T^*$-ordering symbol indicates that derivatives are taken outside the time 
ordering symbol, and the various subscripts give the number of weak fields which
contribute. The analogous Feynman diagrams are shown in Figure~\ref{diagrams}.
\vskip .5cm

\subsubsection{The $D=4$ Result}

The unregulated result \cite{Tsamis:1996qk} can best be understood by considering
how a dimensionally regulated computation of $-i[\mbox{}^{\mu\nu} \Sigma^{\rho
\sigma}](x;x')$ would look. The general forms of the 3-graviton and 4-graviton 
vertices are \cite{Tsamis:1992xa,Tsamis:1996qm},\footnote{Vertices involving
ghosts take the same form as (\ref{vertex3pt}).}
\begin{eqnarray}
\kappa a^{D-2} h \partial h \partial h \qquad & , & \qquad \kappa H a^{D-1} 
h h \partial h \; , \label{vertex3pt} \\
\kappa^2 a^{D-2} h h \partial h \partial h \qquad & , & \qquad \kappa^2 H 
a^{D-1} h h h \partial h \; . \qquad \label{vertex4pt}
\end{eqnarray}
There are a plethora of different index contractions, but contributions to the 
first two (nonlocal) diagrams of Figure~\ref{diagrams} take the general form,
\begin{equation}
\kappa a^{D-2} \times \partial \partial' i\Delta(x;x') \times \partial 
\partial' i\Delta(x;x') \times \kappa {a'}^{D-2} \; , \label{generic3pt}
\end{equation}
with $i\Delta(x;x')$ standing for a ghost or graviton propagator, and the 
understanding that one derivative at each vertex could be replaced by a factor
of $H$ times the appropriate scale factor. Note also that, when an external 
leg happens to be differentiated, then minus the derivative acts on everything. 
On the other hand, the third (4-point) diagram of Figure~\ref{diagrams} is 
local,
\begin{equation}
\kappa^2 a^{D-2} \times \partial \partial' i\Delta(x;x') \times
i\delta^D(x \!-\! x') \; , \label{generic4pt}
\end{equation}
with the same understanding concerning derivatives. The last (counterterm) 
diagram of Figure~\ref{diagrams} is also local,
\begin{equation}
\frac{\kappa^2 a^{D-4}}{D \!-\! 4} \times \partial^2 {\partial'}^2 \times
i\delta^D(x \!-\! x') \; , \label{genericctm}
\end{equation}
with the stipulation that any number of the four derivatives could each be 
replaced by a factor of $H a$.

The gauge for this computation was fixed by adding \cite{Tsamis:1992xa,
Woodard:2004ut},
\begin{equation}
\mathcal{L}_{GF} = -\frac{a^{D-2}}{2} \eta^{\mu\nu} F_{\mu} F_{\nu} 
\;\; , \;\; F_{\mu} = \eta^{\rho\sigma} \Bigl( h_{\mu\rho , \sigma} - 
\frac12 h_{\rho\sigma , \mu} + (D\!-\! 2) a H h_{\mu\rho} \delta^{0}_{~\sigma} 
\Bigr) \; . \label{dSgauge}
\end{equation}
In this gauge the ghost and graviton propagators become sums of constant 
tensor factors multiplied by simple scalar propagators,
\begin{eqnarray}
i\Bigl[\mbox{}_{\mu} \Delta_{\rho}\Bigr](x;x') & = & \overline{\eta}_{\mu\rho}
\times i\Delta_A(x;x') - \delta^0_{~\mu} \delta^0_{~\nu} \times i\Delta_B(x;x')
\; , \label{ghostprop} \\
i\Bigl[\mbox{}_{\mu\nu} \Delta_{\rho\sigma}\Bigr](x;x') & = & \sum_{I=A,B,C} 
\Bigl[\mbox{}_{\mu\nu} T^I_{\rho\sigma}\Bigr] \times i\Delta_I(x;x') \; .
\label{gravprop} 
\end{eqnarray}
The various $[\mbox{}_{\mu\nu} T^I_{\rho\sigma}]$ are,
\begin{eqnarray} 
\Bigl[\mbox{}_{\mu\nu} T^A_{\rho\sigma}\Bigr] = 2 \overline{\eta}_{\mu (\rho}
\overline{\eta}_{\sigma) \nu} - \frac{2}{D\!-\!3} \overline{\eta}_{\mu\nu}
\overline{\eta}_{\rho\sigma} \quad , \quad \Bigl[\mbox{}_{\mu\nu} T^B_{\rho\sigma}
\Bigr] = -4 \delta^0_{~(\mu} \overline{\eta}_{\nu) (\rho} \delta^0_{~ \sigma)} 
\; , \qquad \\
\Bigl[\mbox{}_{\mu\nu} T^C_{\rho\sigma}\Bigr] = \frac{2 E_{\mu\nu} E_{\rho\sigma}}{
(D\!-\!2) (D\!-\!3)} \quad , \quad E_{\mu\nu} \equiv (D\!-\!3) \delta^0_{~\mu}
\delta^0_{~\nu} + \overline{\eta}_{\mu\nu} \; . \qquad 
\end{eqnarray}
Most of the scalar propagators can be expressed using a function $A(y)$ of the
de Sitter length function $y(x;x') \equiv a a' H^2 \Delta x^2$,
\begin{eqnarray}
i\Delta_A(x;x') & = & A(y) + k \ln(a a') \qquad k \equiv 
\frac{H^{D-2}}{(4\pi)^{\frac{D}2}} \frac{\Gamma(D\!-\!1)}{\Gamma(\frac{D}2)} \; , 
\label{DeltaA} \\
i\Delta_B(x;x') & = & B(y) \equiv -\frac{[(4 y \!-\! y^2) A'(y) \!+\! (2 \!-\! y) k]}{
2 (D \!-\! 2)} \; , \label{DeltaB} \\
i\Delta_C(x;x') & = & C(y) \equiv \frac12 (2 \!-\! y) B(y) + \frac{k}{D\!-\!3} \; .
\label{DeltaC}
\end{eqnarray}
The first derivative of $A(y)$ is \cite{Onemli:2002hr,Onemli:2004mb},
\begin{eqnarray}
\lefteqn{A'(y) = -\frac{H^{D-2}}{4 (4 \pi)^{\frac{D}2}} \Biggl\{ \Gamma\Bigl(
\frac{D}2\Bigr) \Bigl( \frac{4}{y}\Bigr)^{\frac{D}2} + \Gamma\Bigl( \frac{D}2 
\!+\! 1\Bigr) \Bigl( \frac{4}{y}\Bigr)^{\frac{D}2 - 1} } \nonumber \\
& & \hspace{2cm} + \sum_{n=0}^{\infty} \Biggl[ \frac{\Gamma(n \!+\! 
\frac{D}2 \!+\! 2)}{\Gamma(n \!+\! 3)} \Bigl( \frac{y}{4}\Bigr)^{n-\frac{D}2 + 2} 
- \frac{\Gamma(n \!+\! D)}{\Gamma(n \!+\! \frac{D}2 \!+\! 1)} \Bigl( 
\frac{y}{4}\Bigr)^{n} \Biggr] \Biggr\} . \qquad \label{Aprime}
\end{eqnarray}
Note that the $y^n$ and $y^{n-\frac{D}2 - 2}$ terms cancel for $D=4$, so they
only contribute when multiplied by a sufficiently singular term.

Divergences occur in the effective field equation (\ref{Einsteineqn}) when
the integration over ${x'}^{\mu}$ carries it to coincidence, ${x'}^{\mu} =
x^{\mu}$. Hence the first two (nonlocal) diagrams of Figure~\ref{diagrams}
can be taken to $D=4$ away from coincidence, which also makes the two local
diagrams vanish. This was done for the unregulated computation
\cite{Tsamis:1996qk}. That computation was tractable because taking $D=4$
simplifies the propagators,
\begin{eqnarray}
i\Bigl[\mbox{}_{\mu} \Delta^{D=4}_{\rho}\Bigr](x;x') & = & \frac1{4\pi^2} \Biggl\{
\frac{\eta_{\mu\rho}}{a a' \Delta x^2} - \frac12 H^2 \ln(H^2 \Delta x^2) 
\overline{\eta}_{\mu\rho} \Biggr\} , \qquad \label{ghostD4} \\
i\Bigl[\mbox{}_{\mu\nu} \Delta^{D=4}_{\rho\sigma}\Bigr](x;x') & = & \frac1{4\pi^2}
\Biggl\{ \frac{(2 \eta_{\mu (\rho} \eta_{\sigma) \nu} \!-\! \eta_{\mu\nu}
\eta_{\rho\sigma})}{a a' \Delta x^2} \nonumber \\
& & \hspace{2cm} - H^2 \ln(H^2 \Delta x^2) \Bigl(\overline{\eta}_{\mu (\rho} 
\overline{\eta}_{\sigma) \nu} \!-\! \overline{\eta}_{\mu\nu} 
\overline{\eta}_{\rho\sigma} \Bigr) \Biggr\} . \qquad \label{gravD4}
\end{eqnarray}
\begin{table}[H]
\setlength{\tabcolsep}{8pt}
\def\arraystretch{1.5}
\centering
\begin{tabular}{|@{\hskip 1mm }c@{\hskip 1mm }||c|}
\hline
$i$ & Coefficient Functions $T^i_{L}(x;x')$ in expression (\ref{TNTLdef}) \\
\hline\hline
$1$ & 
$8 a^2 {a'}^2 H^4 \times [ \frac{4 \Delta \eta^2}{\Delta x^6} + \frac{1}{\Delta x^4}] 
+ 4 a^3 {a'}^3 H^6 \times [ \frac{4 \Delta \eta^4}{\Delta x^6} - \frac{\Delta \eta^2
}{\Delta x^4} + \frac{3}{\Delta x^2}]$ \\
\hline
$2$ & 
$-16 a^2 {a'}^2 H^4 \times [ \frac{4 \Delta \eta^2}{\Delta x^6} + \frac{1}{\Delta x^4}]
- 4 a^3 {a'}^3 H^6 \times [ \frac{8 \Delta \eta^4}{\Delta x^6} + \frac{1}{\Delta x^2}]$ \\
\hline
$3$ & $8 a^3 {a'}^3 H^6 \!\times\! [\frac{\Delta \eta^2}{\Delta x^4} \!-\! 
\frac{2}{\Delta x^2}] \!-\! 4 a^3 {a'}^2 H^5 \!\times\! 
[\frac{4 \Delta \eta^3}{\Delta x^6} \!-\! \frac{\Delta \eta}{\Delta x^4}]$ \\
\hline
$5$ &
$16 a^2 {a'}^2 H^4 \times \frac{\Delta \eta}{\Delta x^4} + 4 a^3 {a'}^2 H^5 \times 
[\frac{2 \Delta \eta^2}{\Delta x^4} + \frac{3}{\Delta x^2}]$ \\
\hline
$7$ &
$-8 a^2 {a'}^2 H^4 \times \frac1{\Delta x^2} - 2 a^3 {a'}^3 H^6 \times 
\frac{\Delta \eta^2}{\Delta x^2} - 2 a^3 {a'}^2 H^5 \times \frac{\Delta \eta}{
\Delta x^2}$ \\
\hline
$9$ &
$-96 a a' H^2 \times [\frac{16 \Delta \eta^4}{\Delta x^{10}} + \frac{12 \Delta \eta^2}{
\Delta x^8} + \frac{1}{\Delta x^6}] - 4 a^2 {a'}^2 H^4 \times [\frac{24 \Delta \eta^4}{
\Delta x^8} + \frac{8 \Delta \eta^2}{\Delta x^6} - \frac{1}{\Delta x^4}]$ \\
\hline
$10$ &
$\!\!96 a a' H^2 \!\times\! [\frac{2 \Delta \eta^3}{\Delta x^8} \!+\! 
\frac{\Delta \eta}{\Delta x^6}] \!+\! 12 a^2 {a'}^2 H^4 \!\times\! 
[\frac{4 \Delta \eta^3}{\Delta x^6} \!+\! \frac{\Delta \eta}{\Delta x^4}]
\!+\! a^3 {a'}^3 H^6 \!\times\! [\frac{8 \Delta \eta^3}{\Delta x^4} \!-\!
\frac{4 \Delta \eta}{\Delta x^2}] \!\!$ \\
$$ & $-8 a^2 a' H^3 \!\times\! [\frac{4 \Delta \eta^2}{\Delta x^6} \!+\! 
\frac1{\Delta x^4}] \!-\! 4 a^3 a'^2 H^5 \!\times\! [\frac{2 \Delta \eta^2}{\Delta x^4} 
\!-\! \frac{1}{\Delta x^2}]$ \\
\hline
$12$ &
$-8 a a' H^2 \!\times\! [\frac{4 \Delta \eta^2}{\Delta x^6} \!+\! \frac{1}{\Delta x^4}]
\!-\! 2 a^2 {a'}^2 H^4 \!\times\! [\frac{6 \Delta \eta^2}{\Delta x^4} \!-\! 
\frac{9}{\Delta x^2}] \!+\! 4 a^3 {a'}^3 H^6 \times \frac{\Delta \eta^2}{\Delta x^2}$ \\
\hline
$13$ & 
$-96 a a' H^2 \times [\frac{16 \Delta \eta^4}{\Delta x^{10}} \!+\! \frac{12 \Delta \eta^2}{
\Delta x^8} \!+\! \frac{1}{\Delta x^6}] + 4 a^2 {a'}^2 H^4 \times [\frac{24 \Delta \eta^4}{
\Delta x^8} \!+\! \frac{56 \Delta \eta^2}{\Delta x^6} \!+\! \frac{11}{\Delta x^4}]$ \\
$$ & $+ 8 a^3 {a'}^3 H^6 \times [ \frac{4 \Delta \eta^4}{\Delta x^6} - 
\frac{2 \Delta \eta^2}{\Delta x^4} + \frac{3}{\Delta x^2}]$ \\
\hline
$14$ & 
$192 a a' H^2 \times [\frac{2 \Delta \eta^3}{\Delta x^8} + \frac{\Delta \eta}{\Delta x^6}]
+ 8 a^2 {a'}^2 H^4 \times [\frac{4 \Delta \eta^3}{\Delta x^6} - \frac{3 \Delta \eta}{
\Delta x^4}]$ \\
$$ & $-16 a^2 a' H^3 \times [\frac{4 \Delta \eta^2}{\Delta x^6} + \frac1{\Delta x^4}]
- 16 a^3 {a'}^2 H^5 \times [\frac{\Delta \eta^2}{\Delta x^4} + \frac{1}{\Delta x^2}]$ \\
\hline
$16$ &
$-8 a a' H^2 \!\times\! [\frac{4 \Delta \eta^2}{\Delta x^6} \!+\! \frac{1}{\Delta x^4}]
\!+\! 2 a^2 {a'}^2 H^4 \!\times\! [-\frac{6 \Delta \eta^2}{\Delta x^4} \!+\! 
\frac1{\Delta x^2}] \!-\! 2 a^3 {a'}^3 H^6 \!\times\! \frac{\Delta \eta^2}{\Delta x^2}$ \\
$$ & $+ 16 a^2 a' H^3 \times \frac{\Delta \eta}{\Delta x^4} + 6 a^3 {a'}^2 H^5 \times
\frac{\Delta \eta}{\Delta x^2}$ \\
\hline
$18$ & 
$-24 a a' H^2 \times [\frac{4 \Delta \eta^2}{\Delta x^6} + \frac{1}{\Delta x^4}]
-2 a^2 {a'}^2 H^4 \times [\frac{6 \Delta \eta^2}{\Delta x^4} + \frac{5}{\Delta x^2}]$ \\
\hline
$19$ & $8 a a' H^2 \times \frac{\Delta \eta}{\Delta x^{4}} + 6 a^2 {a'}^2 H^4
\times \frac{\Delta \eta}{\Delta x^2} - 4 a^2 a' H^3 \times \frac1{\Delta x^2}$ \\
\hline
\end{tabular}
\caption{\footnotesize Each tabulated term must be multiplied by $-
\frac{\kappa^2}{64 \pi^4}$.}
\label{TL}
\end{table}
\noindent Because one of the propagators in the nonlocal diagrams (\ref{generic3pt}) 
might not carry any derivatives, the coefficient functions $T^i(x;x')$ in our
representation (\ref{initialrep}) of the graviton self-energy take the form,
\begin{equation}
T^i(x;x') \equiv T^i_N(x;x') + T^i_L(x;x') \!\times\! \ln(H^2 \Delta x^2) \; .
\label{TNTLdef}
\end{equation}
The coefficient functions $T^i_L(x;x')$ are given in Table~\ref{TL}, and the 
$T^i_N(x;x')$ are given in Table~\ref{TN}. Both are functions of $a$, $a'$, 
$\Delta \eta \equiv \eta - \eta'$ and inverse powers of the Poincar\'e interval 
$\Delta x^2 \equiv \Vert \vec{x} - \vec{x}' \Vert^2 - (\vert \eta - \eta'\vert - 
i \varepsilon)^2$. 
\begin{table}[H]
\setlength{\tabcolsep}{8pt}
\def\arraystretch{1.5}
\centering
\begin{tabular}{|@{\hskip 1mm }c@{\hskip 1mm }||c|}
\hline
$i$ & Coefficient Functions $T^i_{N}(x;x')$ in expression (\ref{TNTLdef}) \\
\hline\hline
$1$ & $\!\!\frac{\frac{736}{5}}{\Delta x^8} \!-\! a a' \!H^2 [\frac{616 \Delta 
\eta^2}{\Delta x^8} \!+\! \frac{\frac{220}{3}}{\Delta x^6}]
\!-\! a^2 {a'}^2\! H^4 [\frac{96 \Delta \eta^4}{\Delta x^8} \!+\! 
\frac{\frac{812}{3} \Delta \eta^2}{\Delta x^6} \!+\! \frac{19}{\Delta x^4}]
\!-\! a^3 {a'}^3 \! H^6 [ \frac{64 \Delta \eta^4}{\Delta x^6} \!+\! 
\frac{22 \Delta \eta^2}{\Delta x^4}]\!\!$ \\
\hline
$2$ & $\!\!\frac{\frac{1952}{5}}{\Delta x^8} \!-\! a a' H^2 [\frac{416 \Delta 
\eta^2}{\Delta x^8} \!+\! \frac{\frac{128}{3}}{\Delta x^6}]
\!-\! a^2 {a'}^2 H^4 [\frac{\frac{112}{3} \Delta \eta^2}{\Delta x^6}
\!-\! \frac{56}{\Delta x^4}] \!+\! a^3 {a'}^3 H^6 [ \frac{32 \Delta \eta^4}{
\Delta x^6} \!+\! \frac{12 \Delta \eta^2}{\Delta x^4}]\!\!$ \\
\hline
$$ & $-184 [\frac{\frac{8}{5} \Delta \eta^2}{\Delta x^{10}} \!+\!
\frac{1}{\Delta x^8}] \!+\! 32 a a' H^2 [ \frac{36 \Delta \eta^4}{
\Delta x^{10}} \!+\! \frac{14 \Delta \eta^2}{\Delta x^8} \!-\! 
\frac{\frac{43}{3}}{\Delta x^6}]$  \\
$3$ & $- a^2 {a'}^2 H^4 [ \frac{288 \Delta \eta^4}{\Delta x^8} \!-\! 
\frac{\frac{1228}{3} \Delta \eta^2}{\Delta x^6} \!+\! \frac{145}{\Delta x^4}]
\!+\! 4 a^3 {a'}^3 H^6 [\frac{16 \Delta \eta^4}{\Delta x^6} \!+\! 
\frac{5 \Delta \eta^2}{\Delta x^4}]$ \\
$$ & $- 8 a H [\frac{232 \Delta \eta^3}{\Delta x^{10}} \!+\! 
\frac{203 \Delta \eta}{\Delta x^8}] \!+\! 4 a^2 a' H^3 
[ \frac{74 \Delta \eta^3}{\Delta x^8} \!-\! \frac{\frac{319}{3} \Delta \eta}{
\Delta x^6}] \!-\! a^3 {a'}^2 H^5 \!\times\! \frac{38 \Delta \eta}{\Delta x^4}$ \\
\hline
$5$ & $\frac{\frac{368}{5} \Delta \eta}{\Delta x^8} \!-\! 32 a a' H^2 
[ \frac{9 \Delta \eta^3}{\Delta x^8} \!-\! \frac{\frac{4}{3} \Delta \eta}{\Delta x^6}] 
\!+\! 4 a^2 {a'}^2 H^4 [ \frac{8 \Delta \eta^3}{\Delta x^6} \!-\!
\frac{\Delta \eta}{\Delta x^4}]$ \\
$$ & $+232 a H [\frac{2 \Delta \eta^2}{\Delta x^8} \!+\! 
\frac{1}{\Delta x^6}] \!-\! 2 a^2 a' H^3
[\frac{\frac{28}{3} \Delta \eta^2}{\Delta x^6} \!-\! \frac{23}{\Delta x^4}] 
\!-\! a^3 {a'}^2 H^5 \!\times\! \frac{12 \Delta \eta^2}{\Delta x^4}$ \\
\hline
$7$ & $-\frac{\frac{92}{15}}{\Delta x^6} \!+\! a a' H^2
[\frac{24 \Delta \eta^2}{\Delta x^6} \!+\! \frac{\frac73}{\Delta x^4}] \!+\!
a^2 {a'}^2 H^4 [\frac{8 \Delta \eta^2}{\Delta x^4} \!-\! 
\frac{3}{\Delta x^2}] \!+\! a^3 {a'}^3 H^6 \!\times\! 
\frac{\Delta \eta^2}{\Delta x^2}$ \\
$$ & $- a H \!\times\! \frac{\frac{116}{3} \Delta \eta}{\Delta x^6} -
a^2 a' H^3 \!\times\! \frac{\frac{23}{3} \Delta \eta}{\Delta x^4} + 
a^3 {a'}^2 H^5 \!\times\! \frac{\Delta \eta}{\Delta x^2}$ \\
\hline
$9$ & $\!\!-16 [\frac{\frac{488}{5} \Delta \eta^2}{\Delta x^{10}} \!+\! 
\frac{61}{\Delta x^8}] \!+\! 16 a a' H^2 [\frac{88 \Delta \eta^4}{
\Delta x^{10}} \!-\! \frac{10 \Delta \eta^2}{\Delta x^8} \!-\! 
\frac{35}{\Delta x^6}]
\!+\! 4 a^2 {a'}^2 H^4 [\frac{8 \Delta \eta^4}{\Delta x^8} \!+\!
\frac{2 \Delta \eta^2}{\Delta x^6} \!-\! \frac{3}{\Delta x^4}]\!\!$ \\
\hline
$10$ & $\frac{\frac{976}{5} \Delta \eta}{\Delta x^8} \!-\! 16 a a' H^2 
[\frac{8 \Delta \eta^3}{\Delta x^8} \!-\! \frac{\frac{23}{3} \Delta 
\eta}{\Delta x^6}] \!-\! 2 a^2 {a'}^2 H^4 [ \frac{\frac{16}{3} \Delta 
\eta^3}{\Delta x^6} \!-\! \frac{23 \Delta \eta}{\Delta x^4}]$ \\
$$ & $+ a^3 {a'}^3 H^6 \!\times\! \frac{4 \Delta \eta^3}{\Delta x^4} \!+\! 
a H \!\times\! \frac{\frac{64}{3}}{\Delta x^6} \!+\! 8 a^2 a' H^3  
[ \frac{\frac{4}{3} \Delta \eta^2}{\Delta x^6} \!-\! \frac{5}{\Delta x^4}] \!-\!
a^3 {a'}^2 H^5 \!\times\! \frac{4 \Delta \eta^2}{\Delta x^4}$ \\  
\hline
$12$ & $-\frac{\frac{488}{15}}{\Delta x^6} \!+\! \frac{32}{3} a a' H^2 
[\frac{\Delta \eta^2}{\Delta x^6} \!-\! \frac1{\Delta x^4}] \!-\! a^2 {a'}^2 H^4 
[\frac{\frac{10}{3} \Delta \eta^2}{\Delta x^4} \!-\! \frac{8}{\Delta x^2}]
\!-\! a^3 {a'}^3 H^6 \!\times\! \frac{2 \Delta \eta^2}{\Delta x^2}$ \\
\hline
$13$ & $16 [\frac{336 \Delta \eta^4}{\Delta x^{12}} \!+\! 
\frac{336 \Delta \eta^2}{\Delta x^{10}} \!+\! \frac{63}{\Delta x^8}] + 4 a a' H^2
[\frac{336 \Delta \eta^4}{\Delta x^{10}} \!+\! \frac{868 \Delta \eta^2}{
\Delta x^8} \!+\! \frac{409}{\Delta x^6}]$ \\ 
$$ & $+ a^2 {a'}^2 H^4 [\frac{424 \Delta \eta^4}{\Delta x^8} \!+\!
\frac{144 \Delta \eta^2}{\Delta x^6} \!+\! \frac{557}{\Delta x^4}] \!-\! 
24 a^3 {a'}^3 H^6 [\frac{4 \Delta \eta^4}{\Delta x^6} \!-\! 
\frac{\Delta \eta^2}{\Delta x^4}]$ \\
\hline
$14$ & $-672 [\frac{\frac{8}{5} \Delta \eta^3}{\Delta x^{10}} \!+\! 
\frac{\Delta \eta}{\Delta x^8}] \!-\! 8 a a' H^2 [\frac{30 \Delta \eta^3}{
\Delta x^8} \!+\! \frac{\frac{107}{3} \Delta \eta}{\Delta x^6}] \!-\! 
4 a^2 {a'}^2 H^4 [\frac{16 \Delta \eta^3}{\Delta x^6} \!-\! 
\frac{7 \Delta \eta}{\Delta x^4}]$ \\
$$ & $-16 a H [\frac{4 \Delta \eta^2}{\Delta x^8} \!+\! \frac{\frac{35}{3}}{
\Delta x^6}] \!-\! 2 a^2 a' H^3 [\frac{\frac{68}{3} \Delta \eta^2}{\Delta x^6} \!+\!
\frac{137}{\Delta x^4}] \!+\! a^3 {a'}^2 H^5 \!\times\! \frac{16 \Delta \eta^2}{
\Delta x^4}$ \\
\hline
$16$ & $4 [\frac{\frac{84}{5} \Delta \eta^2}{\Delta x^8} \!+\! 
\frac{\frac{13}{3}}{\Delta x^6}] \!+\! a a' H^2 [\frac{24 \Delta \eta^2}{
\Delta x^6} \!+\! \frac{\frac{127}{3}}{\Delta x^4}] \!+\! a^2 {a'}^2 H^4 
[\frac{6 \Delta \eta^2}{\Delta x^4} \!+\! \frac{5}{\Delta x^2}]$ \\
$$ & $- a^3 {a'}^3 H^6 \!\times\! \frac{3 \Delta \eta^2}{\Delta x^2} \!-\! a H \!\times\!
\frac{28 \Delta \eta}{\Delta x^6} \!+\! a^2 a' H^3 \!\times\! \frac{\frac{49}{3} 
\Delta \eta}{\Delta x^4} \!+\! a^3 {a'}^2 H^5 \!\times\! \frac{\Delta \eta}{
\Delta x^2}$ \\
\hline
$18$ & $8 [ \frac{\frac{168}{5} \Delta \eta^2}{\Delta x^8} \!+\!
\frac{\frac{29}{3}}{\Delta x^6}] \!+\! 4 a a' H^2
[\frac{12 \Delta \eta^2}{\Delta x^6} \!-\! \frac{23}{\Delta x^4}] \!+\!
a^2 {a'}^2 H^4 \!\times\! \frac{10 \Delta \eta^2}{\Delta x^4}$ \\
\hline
$19$ & $-\frac{\frac{112}{5} \Delta \eta}{\Delta x^6} - 8 a a' H^2 \!\times\!
\frac{\Delta \eta}{\Delta x^4} - a^2 {a'}^2 H^4 \!\times\! \frac{\Delta \eta}{
\Delta x^2} + a H \!\times\! \frac{\frac{50}{3}}{\Delta x^4}$ \\
\hline
$21$ & $\frac{\frac{14}{5}}{\Delta x^4}$ \\
\hline
\end{tabular}
\caption{\footnotesize Each of the tabulated terms must be multiplied by 
$-\frac{\kappa^2}{64 \pi^4}$.}
\label{TN}
\end{table}

\subsubsection{Recovering the Renormalized Result}

In \cite{Tan:2021lza} we presented a 4-step procedure for reconstructing the
dimensionally regulated result for the first two diagrams of Figure~\ref{diagrams}:
\begin{enumerate}
\item{Express each $T^i_L(x;x')$ as a sum of derivatives acting on three 
integrable functions,
\begin{equation}
\frac1{\Delta x^2} \qquad , \qquad \frac{\Delta \eta}{\Delta x^2} \qquad , 
\qquad \frac{\Delta \eta^2}{\Delta x^2} \; ; \label{integrable3}
\end{equation}}
\item{Commute the various derivatives to the left of the multiplicative factor of 
$\ln(H^2 \Delta x^2)$;}
\item{Write the sum of the remainder $\Delta T^i_L(x;x')$ from step 2, and 
$T^i_N(x;x')$, as a sum of derivatives acting on the same integrable functions 
(\ref{integrable3}) and $1/\Delta x^4$; and}
\item{Recognize the factors of $1/\Delta x^4$ from step 3 as the $D=4$ limit of
$1/\Delta x^{2D-4}$, and isolate the ultraviolet divergences on delta functions
which can be absorbed into counterterms.}
\end{enumerate}
\noindent Below we explain the rationale for each step and provide details.
We also implement the various steps on $T^{12}(x;x')$,
\begin{eqnarray}
\lefteqn{T^{12}_L(x;x') = -\frac{\kappa^2 \ln(H^2 \Delta x^2)}{64 \pi^4} \Biggl\{ 
a a' H^2 \Bigl[ -\frac{32 \Delta \eta^2}{\Delta x^6} - \frac{8}{\Delta x^4}\Bigr] } 
\nonumber \\
& & \hspace{3.4cm} + a^2 {a'}^2 H^4 \Bigl[ -\frac{12 \Delta \eta^2}{\Delta x^4}
\!+\! \frac{18}{\Delta x^2}\Bigr] + a^3 {a'}^3 H^6 \Bigl[ 
\frac{4 \Delta \eta^2}{\Delta x^2} \Bigr] \Biggr\} , \qquad \label{T12L1} \\
\lefteqn{T^{12}_N(x;x') = -\frac{\kappa^2}{64 \pi^4} \Biggl\{ -\frac{\frac{488}{15}}{
\Delta x^6} + a a' H \Bigl[ \frac{\frac{32}{3} \Delta \eta^2}{\Delta x^6} - 
\frac{\frac{32}{3}}{\Delta x^6}\Bigr] } \nonumber \\
& & \hspace{3cm} + a^2 {a'}^2 H^4 \Bigl[ -\frac{\frac{10}{3} \Delta \eta^2}{
\Delta x^4} + \frac{8}{\Delta x^2}\Bigr] + a^3 {a'}^3 H^6 \Bigl[-\frac{2 \Delta 
\eta^4}{\Delta x^2} \Bigr] \Biggr\} . \qquad \label{T12N1}
\end{eqnarray}

To understand the rationale behind {\bf Step 1}, note that a single factor 
of $\ln(H^2 \Delta x^2)$ from the propagators (\ref{ghostD4}-\ref{gravD4}) 
can only contribute to one of the $T^i_{L}(x;x')$ if no derivatives act on 
one of the two propagators in (\ref{generic3pt}). In that case all of the 
derivatives must act on the other propagator, and it is this differentiated
propagator, multiplied by the scale factors from the vertices, which appear
in $T^i_{L}(x;x')$. It follows that we can express $T^i_{L}(x;x')$ as a sum
of products of scale factors multiplied by derivatives of the three 
integrable functions (\ref{integrable3}). For example, $T^{12}_{L}(x;x')$
in expression (\ref{T12L1}) can be written as,
\begin{eqnarray}
\lefteqn{T^{12}_L(x;x') = -\frac{\kappa^2 \ln(H^2 \Delta x^2)}{64 \pi^4} 
\Biggl\{ a a' H^2 \!\times\! -4 \partial_0^2 \Bigl( \frac{1}{\Delta x^2} 
\Bigr) } \nonumber \\
& & \hspace{2.8cm} + a^2 {a'}^2 H^4 \Bigl[-6 \partial_0 \Bigl( 
\frac{\Delta \eta}{\Delta x^2} \Bigr) \!+\! \frac{24}{\Delta x^2}\Bigr] 
+ a^3 {a'}^3 H^6 \Bigl[ \frac{4 \Delta \eta^2}{\Delta x^2}\Bigr] \Biggr\} . 
\qquad \label{T12L2}
\end{eqnarray}
The Appendix contains a number of useful identities (\ref{ID1A}-\ref{ID1J}) 
for extracting derivatives. 

{\bf Step 2} consists of commuting the multiplicative factor of $\ln(H^2 
\Delta x^2)$ through the derivatives to multiply the three integrable
functions (\ref{integrable3}). Of course this produces a ``remainder''
$\Delta T^i_{L}(x;x')$ in which derivatives act on the logarithm to produce 
a term like those in $T^i_{N}(x;x')$. For example, carrying out {\bf Step 2}
on expression (\ref{T12L2}) for $T^{12}_{L}(x;x')$ gives,
\begin{eqnarray}
\lefteqn{T^{12}_L(x;x') = -\frac{\kappa^2 }{64 \pi^4} \Biggl\{ a a' H^2 
\!\times\! -4 \partial_0^2 \Bigl( \frac{\ln(H^2 \Delta x^2)}{\Delta x^2} \Bigr) 
+ a^2 {a'}^2 H^4 \Bigl[ -6 \partial_0 } \nonumber \\
& & \hspace{-0.1cm} \times \Bigl( \frac{\Delta \eta \ln(H^2 \Delta x^2)}{\Delta x^2}
\Bigr) \!+\! \frac{24 \ln(H^2 \Delta x^2)}{\Delta x^2} \Bigr] + a^3 {a'}^3 H^6
\!\times\! \frac{4 \Delta \eta^2 \ln(H^2 \Delta x^2)}{\Delta x^2}\Biggr\} 
\qquad \nonumber \\
& & \hspace{2cm} -\frac{\kappa^2}{64 \pi^4} \Biggl\{ a a' H^2 \Bigl[- 
\frac{48 \Delta \eta^2}{\Delta x^6} \!-\! \frac{8}{\Delta x^4} \Bigr] + 
a^2 {a'}^2 H^4 \!\times\! -\frac{12 \Delta \eta^4}{\Delta x^6} \Biggr\} . 
\qquad \label{T12L3}
\end{eqnarray}
Identities (\ref{ID2A}-\ref{ID2J}) in the Appendix facilitate these reductions.
It is useful at this stage to identify six integrable functions, with a factor 
of $2\pi i$ extracted for future convenience,
\begin{eqnarray} 
2 \pi i A_1 \equiv \frac{\ln(H^2 \Delta x^2)}{\Delta x^2} \qquad & , & \qquad
2 \pi i A_2 \equiv \frac{1}{\Delta x^2} \; , \qquad \label{A12def} \\
2 \pi i B_1 \equiv \frac{\Delta \eta\ln(H^2 \Delta x^2)}{\Delta x^2} \qquad & , & 
\qquad 2 \pi i B_2 \equiv \frac{\Delta \eta}{\Delta x^2} \; , \qquad 
\label{B12def} \\
2 \pi i C_1 \equiv \frac{\Delta \eta^2 \ln(H^2 \Delta x^2)}{\Delta x^2} 
\qquad & , & \qquad
2 \pi i C_2 \equiv \frac{\Delta \eta^2}{\Delta x^2} \; . \qquad \label{C12def}
\end{eqnarray}
Hence we can write,
\begin{equation}
T^{12}_L = -\frac{i\kappa^2}{32 \pi^3} \Biggl\{ -4 a a' H^2 \partial_0^2 A_1 
\!-\! 6 a^2 {a'}^2 H^4 \Bigl[\partial_0 B_1 \!-\! 4 A_1\Bigr] \!+\! 4 a^3 {a'}^3
H^6 C_1 \Biggr\} \!+\! \Delta T^2_{L} \; , \label{T12L4}
\end{equation}
where the remainder term is,
\begin{equation}
\Delta T^{12}_L(x;x') = -\frac{\kappa^2}{64 \pi^4} \Biggl\{ a a' H^2 \Bigl[- 
\frac{48 \Delta \eta^2}{\Delta x^6} - \frac{8}{\Delta x^4} \Bigr] + a^2 {a'}^2 
H^4 \!\times\! -\frac{12 \Delta \eta^4}{\Delta x^6} \Biggr\} . \label{T12L5}
\end{equation}
The terms involving $A_1$, $B_1$ and $C_1$ would be ultraviolet finite in
dimensional regularization so it is perfectly valid to leave then in $D=4$.
Results for all the algebraically independent coefficient functions are
given in Table~\ref{NonABC1}.
\begin{table}[H]
\setlength{\tabcolsep}{8pt}
\def\arraystretch{1.5}
\centering
\begin{tabular}{|@{\hskip 1mm }c@{\hskip 1mm }||c|}
\hline
$i$ & Nonlocal Contributions to $i T_{SK}^i(x;x')$ which involve $A_1$, $B_1$ and $C_1$\\
\hline\hline
$1$ & 
$4 a^2 {a'}^2 H^4 \times \partial_0^2 A_1 + 2 a^3 {a'}^3 H^6 \times [ \partial_0^2 C_1
\!-\! 6 \partial_0 B_1 \!+\! 10 A_1]$ \\
\hline
$2$ & $-8 a^2 {a'}^2 H^4 \times \partial_0^2 A_1 \!-\! 4 a^3 {a'}^3 H^6 \times 
[ \partial_0^2 C_1 \!-\! 5 \partial_0 B_1 \!+\! 4 A_1]$ \\
\hline
$3$ & $4 a^3 {a'}^3 H^6 \!\times\! [\partial_0 B_1 \!-\! 5 A_1 ] \!-\! 2 a^3 {a'}^2 H^5 
\!\times\! [\partial_0^2 B_1 \!-\! 4 \partial_0 A_1 ]$ \\
\hline
$5$ &
$8 a^2 {a'}^2 H^4 \times \partial_0 A_1 \!+\! 4 a^3 {a'}^2 H^5 \times [\partial_0 B_1
\!+\! 2 A_1]$ \\
\hline
$7$ &
$-8 a^2 {a'}^2 H^4 \!\times\! A_1 - 2 a^3 {a'}^3 H^6 \!\times\! C_1 - 2 a^3 {a'}^2 H^5 
\!\times\! B_1$ \\
\hline
$9$ & $-4 a a' H^2 \times \partial_0^4 A_1 \!-\! 2 a^2 {a'}^2 H^4 \times [\partial_0^3 B_1
\!-\! 4 \partial_0^2 A_1]$ \\
\hline
$10$ &
$4 a a' H^2 \!\times\! \partial_0^3 A_1 \!+\! 6 a^2 {a'}^2 H^4 \!\times\! [\partial_0^2 B_1
\!-\! 2 \partial_0 A_1 ] $ \\
$$ & $\!\! + a^3 {a'}^3 H^6 \!\!\times\! [ 4 \partial_0 C_1 \!-\! 12 B_1] \!-\! 4 a^2 a' H^3 
\!\!\times\! \partial_0^2 A_1 \!-\! a^3 a'^2 H^5 \!\!\times\! [4 \partial_0 B_1 \!-\! 8 A_1] 
\!\!$ \\
\hline
$12$ &
$-4 a a' H^2 \!\times\! \partial_0^2 A_1 \!-\! 6 a^2 {a'}^2 H^4 \!\times\! [\partial_0 B_1
\!-\! 4 A_1] \!+\! 4 a^3 {a'}^3 H^6 \!\times\! C_1$ \\
\hline
$13$ &
$-4 a a' H^2 \times \partial_0^4 A_1 + 2 a^2 {a'}^2 H^4 \times [ \partial_0^3 B_1
\!+\! 8 \partial_0^2 A_1]$ \\
$$ & $+ 4 a^3 {a'}^3 H^6 \times [ \partial_0^2 C_1 \!-\! 7 \partial_0 B_1 \!+\! 11 A_1 ]$ \\
\hline
$14$ & 
$8 a a' H^2 \times \partial_0^3 A_1 + 4 a^2 {a'}^2 H^4 \times [\partial_0^2 B_1 \!-\! 6 
\partial_0 A_1 ]$ \\
$$ & $-8 a^2 a' H^3 \times \partial_0^2 A_1 \!-\! 8 a^3 {a'}^2 H^5 \times [\partial_0 B_1
\!+\! A_1]$ \\
\hline
$16$ &
$-4 a a' H^2 \!\times\! \partial_0^2 A_1 \!-\! a^2 {a'}^2 H^4 \!\times\! [6 \partial_0 B_1
\!-\! 8 A_1 ] \!-\! 2 a^3 {a'}^3 H^6 \!\times\! C_1$ \\
$$ & $+ 8 a^2 a' H^3 \times \partial_0 A_1 + 6 a^3 {a'}^2 H^5 \!\times\! B_1$ \\
\hline
$18$ & 
$-12 a a' H^2 \!\times\! \partial_0^2 A_1 \!-\! a^2 {a'}^2 H^4 \!\times\! [6 \partial_0 B_1
\!+\! 4 A_1]$ \\
\hline
$19$ & $4 a a' H^2 \times \partial_0 A_1 + 6 a^2 {a'}^2 H^4 \times B_1 - 4 a^2 a' H^3 
\times A_1$ \\
\hline
\end{tabular}
\caption{\footnotesize Each tabulated term must be multiplied by 
$\frac{\kappa^2}{32 \pi^3}$.}
\label{NonABC1}
\end{table}

\begin{table}[H]
\setlength{\tabcolsep}{8pt}
\def\arraystretch{1.5}
\centering
\begin{tabular}{|@{\hskip 1mm }c@{\hskip 1mm }||c|}
\hline
$i$ & Nonlocal Contributions to $i T_{SK}^i(x;x')$ which involve $A_2$, $B_2$ and $C_2$\\
\hline\hline
$1$ & $-a^2 {a'}^2 H^4 [2 \partial_0^3 B_2 \!+\! \frac{95}{6} \partial_0^2 A_2] 
\!-\! a^3 {a'}^3 H^6 [5 \partial_0^2 C_2 \!-\! 12 \partial_0 B_2 \!+\! 2 A_2]$ \\
\hline
$2$ & $-\frac{50}{3} a^2 {a'}^2 H^4 \partial_0^2 A_2 \!-\! a^3 {a'}^3 H^6 
[2 \partial_0^2 C_2 \!-\! 16 \partial_0 B_2 \!+\! 12 A_2]$ \\
\hline
$3$ & $3 a a' H^2 \partial_0^4 A_2 \!-\! a^2 {a'}^2 H^4 [6 \partial_0^3 B_2 \!-\!
\frac{523}{6} \partial_0^2 A_2] \!+\! a^3 {a'}^3 H^6 [8 \partial_0^2 C_2$ \\
$$ & $- 26 \partial_0 B_2 \!+\! 10 A_2] \!+\! \frac{37}{6} a^2 a' H^3 \partial_0^3 
A_2 \!-\! a^3 {a'}^2 H^5 [3 \partial_0^2 B_2 \!+\! 8 \partial_0 A_2]$ \\
\hline
$5$ & $-6 a a' H^2 \partial_0^3 A_2 \!+\! a^2 {a'}^2 H^4 [4 \partial_0^2 B_2 
\!-\! 6 \partial_0 A_2]$ \\
$$ & $- \frac73 a^2 a' H^3 \partial_0^2 A_2 \!-\! 2 a^3 {a'}^2 H^5 
[\partial_0 B_2 \!-\! A_2]$ \\
\hline
$7$ & $3 a a' H^2 \partial_0^2 A_2 \!+\! a^2 {a'}^2 H^4 [4 \partial_0 B_2
\!-\! 7 A_2] \!+\! a^3 {a'}^3 H^6 C_2$ \\
$$ & $-\frac{23}{6} a^2 a' H^3 \partial_0 A_2 \!+\! a^3 {a'}^2 H^5 B_2$ \\
\hline
$9$ & $-\frac{14}{3} a a' H^2 \partial_0^4 A_2 \!-\! a^2 {a'}^2 H^4 
[3 \partial_0^3 B_2 \!-\! 13 \partial_0^2 A_2]$ \\
\hline
$10$ & $\frac{14}{3} a a' H^2 \partial_0^3 A_2 \!+\! a^2 {a'}^2 H^4 
[\frac{23}{3} \partial_0^2 C_2 \!+\! 6 \partial_0 A_2] \!+\! a^3 {a'}^3 H^6
[6 \partial_0 C_2 \!-\! 12 B_2]$ \\
$$ & $-\frac{14}{3} a^2 a' H^3 \partial_0^2 A_2 \!-\! 6 a^3 {a'}^2 H^5
[\partial_0 B_2 \!-\! A_2]$ \\  
\hline
$12$ & $-\frac{14}{3} a a' H^2 \partial_0^2 A_2 \!-\! a^2 {a'}^2 H^4 
[\frac{23}{3} \partial_0 B_2 \!-\! \frac{47}{3} A_2] \!-\! 2 a^3 {a'}^3 H^6
C_2$ \\
\hline
$13$ & $-\frac{29}{6} a a' H^2 \partial_0^4 A_2 \!+\! a^2 {a'}^2 H^4 
[\frac{25}{2} \partial_0^3 B_2 \!-\! 15 \partial_0^2 A_2]$ \\
$$ & $-a^3 {a'}^3 H^6 [6 \partial_0^2 C_2 \!-\! 34 \partial_0 B_2 \!+\! 
22 A_2]$ \\
\hline
$14$ & $\frac{29}{3} a a' H^2 \partial_0^3 A_2 \!-\! a^2 {a'}^2 H^4 
[2 \partial_0^2 B_2 \!-\! 8 \partial_0 A_2] \!-\! \frac{53}{3} a^2 a' H^3 
\partial_0^2 A_2$ \\
\hline
$16$ & $-3 a a' H^2 \partial_0^2 A_2 \!-\! a^2 {a'}^2 H^4 [3 \partial_0 B_2 
\!-\! 8 A_2] \!-\! 3 a^3 {a'}^3 H^6 C_2$ \\
$$ & $+ \frac{97}{6} a^2 a' H^3 \partial_0 A_2 \!+\! a^3 {a'}^2 H^5 B_2$ \\
\hline
$18$ & $-12 a a' H^2 \partial_0^2 A_2 \!-\! a^2 {a'}^2 H^4 
[\partial_0 B_2 \!-\! A_2]$ \\
\hline
$19$ & $-a^2 {a'}^2 H^4 B_2$ \\
\hline
\end{tabular}
\caption{\footnotesize Each of the tabulated terms must be multiplied by 
$\frac{\kappa^2}{32 \pi^3}$.}
\label{NonABC2}
\end{table}

In {\bf Step 3} we first combine $T^i_{N}(x;x')$ with the remainder 
$\Delta T^i_{L}(x;x')$. For our example of $T^{12}(x;x')$ we add 
(\ref{T12N1}) and (\ref{T12L5}),
\begin{eqnarray}
\lefteqn{T^{12}_N(x;x') + \Delta T^{12}_L(x;x') = -\frac{\kappa^2}{64 \pi^4} 
\Biggl\{ -\frac{\frac{488}{15}}{\Delta x^6} + a a' H \Bigl[ -
\frac{\frac{112}{3} \Delta \eta^2}{\Delta x^6} - \frac{\frac{56}{3}}{
\Delta x^4}\Bigr] } \nonumber \\
& & \hspace{3cm} + a^2 {a'}^2 H^4 \Bigl[ -\frac{\frac{46}{3} \Delta \eta^2}{
\Delta x^4} + \frac{8}{\Delta x^2}\Bigr] + a^3 {a'}^3 H^6 \!\times\! -
\frac{2 \Delta \eta^2}{\Delta x^2} \Biggr\} . \qquad \label{T12N2}
\end{eqnarray}
These sums typically contain ultraviolet divergences. If we again employ the
Appendix identities (\ref{ID1A}-\ref{ID1J}) to extract derivatives the result
involves factors of $1/\Delta x^4$ in addition to the three integrable 
functions (\ref{integrable3}). For example, expression (\ref{T12N2}) gives, 
\begin{eqnarray}
\lefteqn{\Bigl[T^{12}_N + \Delta T^{12}_L\Bigr](x;x') = -\frac{\kappa^2}{64 \pi^4} 
\Biggl\{-\partial^2 \Bigl( \frac{\frac{61}{15}}{\Delta x^4} \Bigr) + a a' H^2 
\Bigl[-\partial_0^2 \Bigl( \frac{\frac{14}{3}}{\Delta x^2} \Bigr) - 
\frac{\frac{28}{3}}{\Delta x^4} \Bigr] } \nonumber \\
& & \hspace{2.65cm} + a^2 {a'}^2 H^4 \Bigl[ -\partial_0 \Bigl(\frac{\frac{23}{3} 
\Delta \eta}{\Delta x^2} \Bigr) \!+\! \frac{\frac{47}{3}}{\Delta x^2}\Bigr] 
\!+\! a^3 {a'}^3 H^6 \!\times\! -\frac{2 \Delta \eta^2}{\Delta x^2} \Biggr\} . 
\qquad \label{T12N3}
\end{eqnarray}
The ultraviolet finite factors of $A_2$, $B_2$ and $C_2$ are reported in 
Table~\ref{NonABC2}, whereas we retain the factors of $1/\Delta x^4$ for further 
analysis,
\begin{eqnarray}
\lefteqn{T^{12}_N \!+\! \Delta T^{12}_L = -\frac{i \kappa^2}{32 \pi^3} \Biggl\{-
\frac{14}{3} a a' H^2 \partial_0^2 A_2 \!+\! a^2 {a'}^2 H^4 \Bigl[-\frac{23}{3}
\partial_0 B_2 \!+\! \frac{47}{3} A_2\Bigr] } \nonumber \\
& & \hspace{2.9cm} -2 a^3 {a'}^3 H^6 C_2 \Biggr\} -\frac{\kappa^2}{64 \pi^4} 
\Biggl\{ -\partial^2 \Bigl( \frac{\frac{61}{15}}{\Delta x^4} \Bigr) \!-\! 
a a' H^2 \frac{\frac{28}{3}}{\Delta x^4} \Biggr\} . \qquad \label{T12N4}
\end{eqnarray}

In {\bf Step 4} we isolate the logarithmic ultraviolet divergence implicit in
the factors of $1/\Delta x^4$ produced by {\bf Step 3}. We first note that 
factors of $1/\Delta x^4$ would appear as $1/\Delta x^{2D-4}$ had dimensional
regularization been retained. Extracting a d'Alembertian from this uncovers an
explicit factor of $1/(D-4)$,
\begin{equation}
\frac1{\Delta x^4} \longrightarrow \frac1{\Delta x^{2D-4}} = \frac{\partial^2}{
2 (D\!-\!3) (D\!-\!4)} \Bigl[\frac1{\Delta x^{2D-6}} \Bigr] \; . \label{newID1}
\end{equation}
The ultraviolet divergence is localized by adding a term proportional to the
flat space background massless propagator equation \cite{Onemli:2002hr,
Onemli:2004mb},
\begin{eqnarray}
\lefteqn{\frac1{\Delta x^4} \longrightarrow \frac{\partial^2}{
2 (D\!-\!3) (D\!-\!4)} \Bigl[\frac1{\Delta x^{2D-6}} \Bigr] } \nonumber \\
& & \hspace{0cm} = \frac{\partial^2}{2 (D\!-\!3) (D\!-\!4)} \Bigl[\frac1{\Delta 
x^{2D-6}} - \frac{\mu^{D-4}}{\Delta x^{D-2}} \Bigr] + \frac{\mu^{D-4} 4 \pi^{\frac{D}2} 
i \delta^D(x \!-\! x')}{2 (D\!-\!3) (D\!-\!4) \Gamma(\frac{D}2 \!-\! 1)} \; . \qquad
\label{newID2}
\end{eqnarray}
The nonlocal part of (\ref{newID2}) is both integrable and finite for $D=4$.
We can take the unregulated limit of the nonlocal part of (\ref{newID2}),
\begin{equation}
\frac{\partial^2}{2 (D\!-\!3) (D\!-\!4)} \Bigl[\frac1{\Delta x^{2D-6}} - 
\frac{\mu^{D-4}}{\Delta x^{D-2}} \Bigr] \longrightarrow -\frac{\partial^2}{4} 
\Bigl[ \frac{\ln(\mu^2 \Delta x^2)}{\Delta x^2}\Bigr] \equiv -\frac{\partial^2}{4}
\Bigl[ 2\pi i A_3\Bigr] \; . \label{A3def}
\end{equation}
These ultraviolet finite terms are given in Table~\ref{NonA3}.
\begin{table}[H]
\setlength{\tabcolsep}{8pt}
\def\arraystretch{1.5}
\centering
\begin{tabular}{|@{\hskip 1mm }c@{\hskip 1mm }||c|}
\hline
$i$ & Nonlocal Contributions to $i T_{SK}^i(x;x')$ which involve $A_3$\\
\hline\hline
$1$ & $-\frac{23}{120} \partial^6 A_3 \!+\! a a' H^2 [\frac{77}{12}
\partial_0^2 \!-\! \frac{11}{12} \partial^2] \partial^2 A_3 \!-\! 
\frac{49}{6} a^2 {a'}^2 H^4 \partial^2 A_3$ \\
\hline
$2$ & $-\frac{61}{120} \partial^6 A_3 \!+\! a a' H^2  
[\frac{13}{3} \partial_0^2 \!-\! \frac{5}{6} \partial^2] \partial^2 A_3
\!-\! \frac{55}{3} a^2 {a'}^2 H^4 \partial^2 A_3$ \\
\hline
$3$ & $\frac{23}{120} \nabla^2 \partial^4 A_3 \!+\! a a' H^2 [\frac{52}{12}
\partial_0^2 \!+\! \frac{173}{12} \partial^2] \partial^2 A_3 \!+\! \frac{425}{6} 
a^2 {a'}^2 H^4 \partial^2 A_3$ \\
$$ & $+ a H [\frac{29}{12} \partial_0^3 \!+\! \frac{29}{6} \partial_0 \partial^2] 
\partial^2 A_3 \!+\! \frac{215}{6} a^2 a' H^3 \partial_0 \partial^2 A_3$ \\
\hline
$5$ & $-\frac{23}{60} \partial_0 \partial^4 A_3 \!-\! \frac{35}{3} a a' H^2 
\partial_0 \partial^2 A_3 \!-\! \frac{29}{6} a H [\partial_0^2 \!+\! \partial^2] 
\partial^2 A_3 \!-\! \frac{38}{3} a^2 a' H^3 \partial^2 A_3$ \\
\hline
$7$ & $\frac{23}{120} \partial^4 A_3 \!+\! \frac{11}{12} a a' H^2 \partial^2 A_3
\!+\! \frac{29}{12} a H \partial_0 \partial^2 A_3$ \\
\hline
$9$ & $\frac{61}{60} \nabla^2 \partial^4 A_3 \!+\! a a' H^2 [\frac{29}{3} 
\partial_0^2 \!+\! \frac{41}{3} \partial^2] \partial^2 A_3 \!+\! 4 a^2 {a'}^2 
H^4 \partial^2 A_3$ \\
\hline
$10$ & $-\frac{61}{60} \partial_0 \partial^4 A_3 \!-\! \frac{29}{3} a a' H^2 
\partial_0 \partial^2 A_3 \!-\! \frac{2}{3} a H \partial^4 A_3 \!+\! \frac{29}{3} 
a^2 a' H^3 \partial^2 A_3$ \\ 
\hline
$12$ & $\frac{61}{60} \partial^4 A_3 \!+\! \frac{7}{3} a a' H^2 \partial^2 A_3$ \\
\hline
$13$ & $-\frac{7}{10} \nabla^4 \partial^2 \!A_3 \!-\! a a' H^2 [\frac{86}{3}
\partial_0^2 \!+\! \frac{431}{12} \partial^2] \partial^2 \!A_3 \!-\! 
139 a^2 {a'}^2 H^4 \partial^2 \!A_3\!\!$ \\
\hline
$14$ & $\frac{7}{5} \nabla^2 \partial_0 \partial^2 A_3 \!+\! \frac{43}{3} a a' 
H^2 \partial_0 \partial^2 A_3 \!+\! a H [\frac{2}{3} \partial_0^2 \!+\! 
\frac{11}{2} \partial^2] \partial^2 \!A_3 \!+\! \frac{191}{3} a^2 a' H^3 
\partial^2 A_3$ \\
\hline
$16$ & $-[\frac{7}{10} \partial_0^2 \!+\! \frac{23}{120} \partial^2] \partial^2
A_3 \!-\! \frac{121}{12} a a' H^2 \partial^2 A_3 \!+\! \frac{7}{4} a H 
\partial_0 \partial^2 A_3$ \\
\hline
$18$ & $-[\frac{14}{5} \partial_0^2 \!+\! \frac{61}{60} \partial^2] \partial^2 
A_3 \!+\! 23 a a' H^2 \partial^2 A_3$ \\
\hline
$19$ & $\frac{7}{5} \partial_0 \partial^2 A_3 \!-\! \frac{25}{6} a H \partial^2
A_3$ \\
\hline
$21$ & $-\frac{7}{10} \partial^2 A_3$ \\
\hline
\end{tabular}
\caption{\footnotesize Each of the tabulated terms must be multiplied by 
$\frac{\kappa^2}{32 \pi^3}$.}
\label{NonA3}
\end{table}

It remains to renormalize the local divergence in expression (\ref{newID2}).
This turns out to always produce a finite local term proportional to $\ln(a)$.
It arises from the incomplete cancellation between primitive divergences like
(\ref{newID2}) and counterterms, which contain an extra factor of $a^{D-4}$
from the measure,
\begin{eqnarray}
\lefteqn{\frac{\mu^{D-4} 4 \pi^{\frac{D}2} i \delta^D(x \!-\! x')}{2 (D\!-\!3) 
(D\!-\!4) \Gamma(\frac{D}2 \!-\! 1)} - \frac{a^{D-4} \mu^{D-4} 4 \pi^{\frac{D}2}
i \delta^D(x \!-\! x')}{2 (D\!-\!3) (D\!-\!4) \Gamma(\frac{D}2 \!-\! 1)} }
\nonumber \\
& & \hspace{7cm} \longrightarrow - 2\pi^2 i \!\times\! \ln(a) 
\delta^4(x \!-\! x') \; . \qquad \label{renorm} 
\end{eqnarray}
These local terms are reported in Table~\ref{Local}.
\begin{table}[H]
\setlength{\tabcolsep}{8pt}
\def\arraystretch{1.5}
\centering
\begin{tabular}{|@{\hskip 1mm }c@{\hskip 1mm }||c|}
\hline
$i$ & Local Contributions to $i T_{SK}^i(x;x')$ \\
\hline\hline
$1$ & $-\frac{23}{30} \partial^4 \delta^4(x\!-\!x') \!+\! a a' H^2 [\frac{77}{3} 
\partial_0^2 \!-\! \frac{11}{3} \partial^2] \delta^4(x \!-\! x') \!-\! 
\frac{98}{3} a^2 {a'}^2 H^4 \delta^4(x \!-\! x')$ \\
\hline
$2$ & $-\frac{61}{30} \partial^4 \delta^4(x \!-\! x') \!+\! a a' H^2  
[\frac{52}{3} \partial_0^2 \!-\! \frac{10}{3} \partial^2] \delta^4(x \!-\! x')
\!-\! \frac{220}{3} a^2 {a'}^2 H^4 \delta^4(x \!-\! x')$ \\
\hline
$3$ & $\frac{23}{30} \nabla^2 \partial^2 \delta^4(x \!-\! x') \!+\! a a' H^2 
[\frac{52}{3} \partial_0^2 \!+\! \frac{173}{3} \partial^2] \delta^4(x \!-\! x') 
\!+\! \frac{850}{3} a^2 {a'}^2 H^4$ \\
$$ & $\times \delta^4(x \!-\! x') \!+\! a H [\frac{29}{3} 
\partial_0^3 \!+\! \frac{58}{3} \partial_0 \partial^2] \delta^4(x \!-\! x') 
\!+\! \frac{430}{3} a^2 a' H^3 \partial_0 \delta^4(x \!-\! x')$ \\
\hline
$5$ & $-\frac{23}{15} \partial_0 \partial^2 \delta^4(x \!-\! x') \!-\! 
\frac{140}{3} a a' H^2 \partial_0 \delta^4(x \!-\! x')$ \\
$$ & $ - \frac{58}{3} a H \nabla^2 \delta^4(x \!-\! x') 
\!-\! \frac{152}{3} a^2 a' H^3 \delta^4(x \!-\! x')$ \\
\hline
$7$ & $\frac{23}{30} \partial^2 \delta^4(x \!-\! x') \!+\! \frac{11}{3} a a' H^2
\delta^4(x \!-\! x') \!+\! \frac{29}{3} a H \partial_0 \delta^4(x \!-\! x')$ \\
\hline
$9$ & $\frac{61}{15} \nabla^2 \partial^2 \delta^4(x \!-\! x') \!+\! a a' H^2 
[\frac{116}{3} \partial_0^2 \!+\! \frac{164}{3} \partial^2] \delta^4(x \!-\! x')
\!+\! 16 a^2 {a'}^2 H^4 \delta^4(x \!-\! x')$ \\
\hline
$10$ & $-\frac{61}{15} \partial_0 \partial^2 \delta^4(x \!-\! x') \!-\! 
\frac{116}{3} a a' H^2 \partial_0 \delta^4(x \!-\! x') \!-\! \frac{8}{3} a H 
\partial^2 \delta^4(x \!-\! x')$ \\
$$ & $+ \frac{116}{3} a^2 a' H^3 \delta^4(x \!-\! x')$ \\ 
\hline
$12$ & $\frac{61}{15} \partial^2 \delta^4(x \!-\! x') \!+\! \frac{28}{3} a a' H^2 
\delta^4(x \!-\! x')$ \\
\hline
$13$ & $-\frac{14}{5} \nabla^4 \delta^4(x \!-\! x') \!-\! a a' H^2 [\frac{344}{3}
\partial_0^2 \!+\! \frac{431}{3} \partial^2] \delta^4(x \!-\! x') \!-\! 556 a^2 
{a'}^2 H^4 \delta^4(x \!-\! x')$ \\
\hline
$14$ & $\frac{28}{5} \nabla^2 \partial_0 \delta^4(x \!-\! x') \!+\! \frac{172}{3} 
a a' H^2 \partial_0 \delta^4(x \!-\! x')$ \\
$$ & $+a H [\frac{8}{3} \partial_0^2 \!+\! 22 \partial^2] \delta^4(x \!-\! x')
\!+\! \frac{764}{3} a^2 a' H^3 \delta^4(x \!-\! x')$ \\
\hline
$16$ & $-[\frac{14}{5} \partial_0^2 \!+\! \frac{23}{30} \partial^2] 
\delta^4(x \!-\! x') \!-\! \frac{121}{3} a a' H^2 \delta^4(x \!-\! x') \!+\!
7 a H \partial_0 \delta^4(x \!-\! x')$ \\
\hline
$18$ & $-[\frac{56}{5} \partial_0^2 \!+\! \frac{61}{15} \partial^2] 
\delta^4(x \!-\! x') \!+\! 92 a a' H^2 \delta^4(x \!-\! x')$ \\
\hline
$19$ & $\frac{28}{5} \partial_0 \delta^4(x \!-\! x') \!-\! \frac{50}{3} a H
\delta^4(x \!-\! x')$ \\
\hline
$21$ & $-\frac{14}{5} \delta^4(x \!-\! x')$ \\
\hline
\end{tabular}
\caption{\footnotesize Each of the tabulated terms must be multiplied by 
$\frac{\kappa^2 \ln(a)}{32 \pi^2}$.}
\label{Local}
\end{table}
\noindent To see that primitive divergences are free of $D$-dependent scale
factors, note first that the two nonlocal diagrams of Figure~\ref{diagrams},
corresponding to the generic expression (\ref{generic3pt}), acquire a factor of 
$(a a')^{D-2}$ from the two 3-point vertices. The $D$-dependence of these vertex 
scale factors is cancelled by scale factors from the two propagators. The most 
singular part of each propagator is,
\begin{equation}
\frac{H^{D-2} \Gamma(\frac{D}2 \!-\! 1)}{(4 \pi)^{\frac{D}2}} \Bigl(
\frac{4}{y} \Bigr)^{\frac{D}{2}-1} = \frac{\Gamma(\frac{D}2 \!-\! 1)}{4 
\pi^{\frac{D}2}} \Bigl( \frac{1}{a a' \Delta x^2}\Bigr)^{\frac{D}2-1} \; .
\end{equation}
Less singular terms differ among the various propagators, but their scale 
factors all have the form $(a a')^{1 - \frac{D}2} \times (a a')^N$ 
necessary to cancel the $D$-dependence of the vertex scale factors.

\subsubsection{The Schwinger-Keldysh Result}

Even though the graviton field is Hermitian, the nonlocal factors 
(\ref{A3def}) and (\ref{A12def}-\ref{C12def}) are neither real nor causal 
because the Feynman diagrams from which they derive are in-out matrix 
elements rather than expectation values. We can derive true expectation 
values using the Schwinger-Keldysh formalism \cite{Schwinger:1960qe,
Mahanthappa:1962ex,Bakshi:1962dv,Bakshi:1963bn,Keldysh:1964ud} which is a 
diagrammatic technique that is almost as simple as the Feynman rules. These
expectation values obey effective field equations that are real and causal,
albeit nonlocal \cite{Chou:1984es,Jordan:1986ug,Calzetta:1986ey}.

There is no point to deriving the rules for converting the 1PI $N$-point
functions such as $-i [\mbox{}^{\mu\nu} \Sigma^{\rho\sigma}](x;x')$ from 
in-out amplitudes to the Schwinger-Keldysh formalism. We merely list the
rules \cite{Ford:2004wc}:
\begin{itemize}
\item{Spacetime points carry a $\pm$ polarity.}
\item{Because propagators have two points, each with two polarities, 
there are four Schwinger-Keldysh propagators $i\Delta_{\pm \pm}(x;x')$. The
$++$ case is just the Feynman propagator, whereas the $--$ case is its 
conjugate. The $-+$ propagator is the free expectation value of the field 
at $x^{\mu}$ times the field at ${x'}^{\mu}$, and the $+-$ propagator is 
the free expectation value of the reverse-ordered product.}
\item{Each vertex has a $\pm$ polarity. The $+$ vertices are the same as 
those of the in-out formalism while the $-$ vertices are complex conjugates.}
\item{Every in-out 1PI $N$-point function gives rise to $2^N$ $N$-point 
functions in the Schwinger-Keldysh formalism.} 
\item{The factor of $[\mbox{}^{\mu\nu} \Sigma^{\rho\sigma}](x;x')$ in the
linearized quantum Einstein equation (\ref{Einsteineqn}) is replaced by the
sum of $[\mbox{}^{\mu\nu} \Sigma^{\rho\sigma}_{++}](x;x')$, which is the same 
as the in-out result, and $[\mbox{}^{\mu\nu} \Sigma^{\rho\sigma}_{+-}](x;x')$.}
\item{On our simple background (\ref{geometry}), one can infer the result for
$[\mbox{}^{\mu\nu} \Sigma^{\rho\sigma}_{+-}](x;x')$ from that for 
$[\mbox{}^{\mu\nu} \Sigma^{\rho\sigma}](x;x')$ by dropping all the local
contributions of Table~\ref{Local}, multiplying the nonlocal terms by $-1$,
and converting the coordinate interval $\Delta x^2$ from
\begin{equation}
\Delta x^2_{++}(x;x') \equiv \Bigl\Vert \vec{x} - \vec{x}' \Bigr\Vert^2 -
\Bigl( \vert \eta \!-\! \eta'\vert - i \varepsilon \Bigr)^2 \; , 
\label{Dx++}
\end{equation}
to
\begin{equation}
\Delta x^2_{+-}(x;x') \equiv \Bigl\Vert \vec{x} - \vec{x}' \Bigr\Vert^2 -
\Bigl( \eta \!-\! \eta' + i \varepsilon \Bigr)^2 \; . 
\label{Dx+-}
\end{equation}} 
\end{itemize}

Implementing these rules is straightforward. First, recall that the only 
dependence on the coordinate interval $\Delta x^2$ in the nonlocal results 
of Tables~\ref{NonABC1}, \ref{NonABC2} and \ref{NonA3} comes through the 
integrable functions $A_{1-3}$, $B_{12}$ and $C_{1-2}$, which were defined 
in expressions (\ref{A12def}-\ref{C12def}) and (\ref{A3def}). We can 
eliminate the factors of $1/\Delta x^2$ using identities 
(\ref{ID3A}-\ref{ID3I}) of the Appendix. For example, the $++$ and $+-$
versions of $2 \pi i \times A_1$ are,
\begin{equation}
2\pi i \!\times\! A_1 = \frac{\ln(H^2 \Delta x^2_{+\pm})}{\Delta x^2_{+\pm}}
= \frac{\partial^2}{8} \Bigl[ \ln^2(H^2 \Delta x^2_{+\pm}) \!-\! 2 \ln(H^2
\Delta x^2_{+\pm}) \Bigr] \; .
\end{equation}
Because the scale factors and derivatives are identical in the $++$ and
$+-$ contributions, we just need to consider differences of logarithms,
\begin{eqnarray}
\ln(H^2 \Delta x^2_{++}) - \ln(H^2 \Delta x^2_{+-}) & = & 2\pi i \!\times\!
\theta(\Delta \eta \!-\! r) \; , \\
\ln^2(H^2 \Delta x^2_{++}) - \ln^2(H^2 \Delta x^2_{+-}) & = & 4\pi i \!\times\!
\theta(\Delta \eta \!-\! r) \ln[H^2 (\Delta \eta^2 \!-\! r^2)] \; , \qquad 
\end{eqnarray} 
where $r \equiv \Vert \vec{x} - \vec{x}'\Vert$. For example, the factors of
$A_1$ on Table~\ref{NonABC1} have the Schwinger-Keldysh correspondence,
\begin{equation}
A_1 \longrightarrow +\frac{\partial^2}{4} \Biggl\{ \theta(\Delta \eta \!-\! r) 
\Bigl[ \ln[ H^2 (\Delta \eta^2 \!-\! r^2)] \!-\! 1\Bigr] \Biggr\} \; .
\end{equation}
Identities (\ref{ID4A}-\ref{ID4G}) in the Appendix give the reductions needed
for any of the integrable functions $A_{1-3}$, $B_{1-2}$ and $C_{1-2}$.

\subsection{The 4-Point Contribution}

The previous discussion concerned the two nonlocal diagrams of 
Figure~\ref{diagrams}, and the local counterterms needed to renormalize them.
There are also finite local contributions from the 3rd diagram. It derives 
from the 42 4-graviton interactions given in equation (4.1) of 
\cite{Tsamis:1996qm}. One connects two of the graviton fields to the external
legs and then replaces the remaining two fields by graviton propagator. The 
procedure is tedious and we shall content ourselves with simply sketching it
and giving the final result. 

As an example we reduce the first of the 42 interactions, 
\begin{equation}
S_1 \equiv \frac{\kappa^2}{32} \!\int\!\! d^Dx \, a^{D-2} h^2 h_{,\theta} 
h^{,\theta} \; ,
\end{equation} 
where a comma denotes differentiation and the trace of the graviton field is
$h \equiv h^{\alpha}_{~\alpha} \equiv \eta^{\alpha\beta} h_{\alpha\beta}$. 
We first take variational derivatives of the action integral with respect to 
$h_{\mu\nu}(x)$ and $h_{\rho\sigma}(x')$ as in expression (\ref{operatorexpr}),
\begin{eqnarray}
\lefteqn{\frac{i \delta^2 S_1}{\delta h_{\mu\nu}(x) \delta h_{\rho\sigma}(x')} =
\frac{\kappa^2}{32} \eta^{\mu\nu} \eta^{\rho\sigma} \Biggl\{ -\partial_{\theta}
\Bigl[ 2 a^{D-2} h^{\alpha}_{~\alpha}(x) h^{\beta}_{~\beta}(x) \partial^{\theta}
i \delta^D(x\!-\!x') \Bigr] } \nonumber \\
& & \hspace{0cm} + 4 a^{D-2} h^{\alpha}_{~\alpha}(x) h^{\beta}_{~\beta , 
\theta}(x) \partial^{\theta} i \delta^D(x \!-\! x') \!-\! 4 \partial^{\theta} 
\Bigl[ a^{D-2} h^{\alpha}_{~\alpha}(x) h^{\beta}_{~\beta , \theta}(x) 
i \delta^D(x \!-\! x') \Bigr] \nonumber \\
& & \hspace{5.5cm} + 2 a^{D-2} h^{\alpha}_{~\alpha , \theta}(x)
h^{\beta}_{~\beta ,\theta}(x) \Big{]} i \delta^D(x\!-\!x') \Biggr\} . \qquad 
\end{eqnarray}
Now compute the expectation value of the $T^*$-ordered product, which amounts 
to replacing the remaining two graviton fields of each term by the appropriate 
coincident (and sometimes differentiated) propagator,
\begin{eqnarray}
\lefteqn{\Bigl\langle \Omega \Bigl\vert T^*\Bigl[\frac{i \delta^2 S_1}{
\delta h_{\mu\nu}(x) \delta h_{\rho\sigma}(x')} \Bigr] \Bigr\vert \Omega 
\Bigr\rangle = \frac{\kappa^2}{32} \eta^{\mu\nu} \eta^{\rho\sigma} \Biggl\{ 
-\partial_{\theta} \Biggl[ 2 a^{D-2} \!\times\! i\!\Bigl[\mbox{}^{\alpha}_{~\alpha} 
\Delta^{\beta}_{~\beta}\Bigr]\!(x;x) } \nonumber \\
& & \hspace{-0.7cm} \times \partial^{\theta} i \delta^D(x\!-\!x') \Biggr] \!+\! 
4 a^{D-2} \!\times\! \partial'_{\theta} i\!\Bigl[\mbox{}^{\alpha}_{~\alpha} 
\Delta^{\beta}_{~\beta}\Bigr]\!(x;x')\Bigl\vert_{x'=x} \!\!\!\!\!\!\!\!\times 
\partial^{\theta} i \delta^D\!(x \!-\! x') \!-\! 4 \partial^{\theta} \Biggl[ 
a^{D-2}  \nonumber \\
& & \hspace{-0.7cm} \times i\delta^D\!(x \!-\! x') \!\times\! \partial'_{\theta} 
i\!\Bigl[\mbox{}^{\alpha}_{~\alpha} \Delta^{\beta}_{~\beta} \Bigr]\!(x;x') \!\Biggr] 
\!\!+\! 2 a^{D-2} i \delta^D\!(x\!-\!x') \!\times\! \partial_{\theta} 
\partial'^{\theta} i\!\Bigl[\mbox{}^{\alpha}_{~\alpha} \Delta^{\beta}_{~\beta}
\Bigr]\!(x;x') \!\Biggr\} . \qquad
\end{eqnarray}
Finally, we express the tensor structure using the 21 basis tensors of 
Table~\ref{Tbasis},
\begin{eqnarray}
\eta^{\mu\nu} \eta^{\rho\sigma} & = & \Bigl( \overline{\eta}^{\mu\nu} \!-\!
\delta^{\mu}_{~0} \delta^{\nu}_{~0} \Bigr) \Bigl(\overline{\eta}^{\rho\sigma}
\!-\! \delta^{\rho}_{~0} \delta^{\sigma}_{~0}\Bigr) \; , \\
& = & \Bigl[\mbox{}^{\mu\nu} \mathcal{D}_1^{\rho\sigma}\Bigr] \!-\!
\Bigl[\mbox{}^{\mu\nu} \mathcal{D}_3^{\rho\sigma}\Bigr] \!-\!
\Bigl[\mbox{}^{\mu\nu} \mathcal{D}_4^{\rho\sigma}\Bigr] \!+\!
\Bigl[\mbox{}^{\mu\nu} \mathcal{D}_{13}^{\rho\sigma}\Bigr] \; . \qquad 
\end{eqnarray}

The coincidence limits of the three propagators which appear in the graviton
propagator (\ref{gravprop}) are,
\begin{eqnarray} 
i\Delta_A(x;x) = k \Bigl[-\pi {\rm cot}\Bigl( \frac{\pi D}{2}\Bigr) \!+\! 2
\ln(a) \Bigr] \quad & , & \quad i\Delta_B(x;x) = -\frac{k}{D\!-\!2} \; , 
\qquad \\
i\Delta_C(x;x) = \frac{k}{(D\!-\!2)(D\!-\!3)} \quad & , & \quad k \equiv 
\frac{H^{D-2}}{(4\pi)^{\frac{D}2}} \frac{\Gamma(D\!-\!1)}{\Gamma(\frac{D}2)}
\; . \qquad 
\end{eqnarray}
Note that only the undifferentiated $A$-type propagator is ultraviolet
divergent in dimensional regularization. The undifferentiated $A$-type propagator
is also the only way to get a factor of $\ln(a)$. First derivatives of coincident 
propagators are all finite,
\begin{equation}
\partial_{\alpha} i\Delta_A(x;x') \Bigl\vert_{x'=x} = a H k \delta^0_{~\alpha}
\;\; , \;\; \partial_{\alpha} i\Delta_B(x;x') \Bigl\vert_{x'=x} = 0 =
\partial_{\alpha} i\Delta_C(x;x') \Bigl\vert_{x'=x} \; .
\end{equation}
Mixed second derivatives are also finite,
\begin{eqnarray}
\partial_{\alpha} \partial'_{\beta} i\Delta_A(x;x') \Bigl\vert_{x'=x} & = &
-\Bigl( \frac{D\!-\!1}{D} \Bigr) k H^2 g_{\alpha\beta} \; , \\
\partial_{\alpha} \partial'_{\beta} i\Delta_B(x;x') \Bigl\vert_{x'=x} & = &
\frac{1}{D} k H^2 g_{\alpha\beta} \; , \\
\partial_{\alpha} \partial'_{\beta} i\Delta_C(x;x') \Bigl\vert_{x'=x} & = &
-\frac{2}{D (D \!-\! 2)} k H^2 g_{\alpha\beta} \; .
\end{eqnarray}

\begin{table}[H]
\setlength{\tabcolsep}{8pt}
\def\arraystretch{1.5}
\centering
\begin{tabular}{|@{\hskip 1mm }c@{\hskip 1mm }||c|}
\hline
$i$ & Nonzero contributions to $i T_{SK}^i(x;x')$ from the 4-point diagram \\
\hline\hline
$1$ & $-8 a^2 H^2 (\partial_0 \!+\! 2 a H) \partial_0 \delta^4(x \!-\! x')$ \\
\hline
$2$ & $8 a^2 H^2 (\partial_0 \!+\! 2 a H) \partial_0 \delta^4(x \!-\! x')$ \\
\hline
$3$ & $- 8 a^2 H^2 [\nabla^2 \!+\! a H \partial_0 \!+\! 3 a^2 H^2] 
\delta^4(x \!-\! x')$ \\
\hline
$5$ & $16 a^2 H^2 (\partial_0 \!+\! 2 a H) \delta^4(x \!-\! x')$ \\
\hline
$9$ & $16 a^2 H^2 \nabla^2 \delta^4(x \!-\! x')\!\!$ \\
\hline
$10$ & $-16 a^2 H^2 \partial_0 \delta^4(x \!-\! x')$ \\
\hline
$13$ & $-72 a^4 H^4 \delta^4(x \!-\! x')$ \\
\hline
$14$ & $-16 a^3 H^3 \delta^4(x \!-\! x')$ \\
\hline
$16$ & $8 a^2 H^2 \delta^4(x \!-\! x')$ \\
\hline
$18$ & $-16 a^2 H^2 \delta^4(x \!-\! x')$ \\
\hline
\end{tabular}
\caption{\footnotesize Each of the tabulated terms must be multiplied by 
$\frac{\kappa^2 \ln(a)}{32 \pi^2}$.}
\label{Local4pt}
\end{table}

Note that all primitive contributions have factors of $a^{D-2}$, $a^{D-1}$
or $a^D$. The counterterms which absorb ultraviolet divergences possess 
the very same dependence on $a$ so renormalization engenders no finite factors 
of $\ln(a)$ the way it did for the nonlocal diagrams of expression 
(\ref{renorm}). It does produce factors of $\ln(H/\mu)$ but we report only the
$\ln(a)$ contributions in Table~\ref{Local4pt}.

\subsection{Anomalous Local Contributions}

Our result for the renormalized self-energy consists of the local 
contributions, collected in Tables~\ref{Local} and \ref{Local4pt}, plus the
nonlocal contributions of Tables~\ref{NonABC1}, \ref{NonABC2} and \ref{NonA3}.
The nonlocal contributions obey the Ward identity, just as did the 
noncoincident, $D=4$ result \cite{Tsamis:1996qk} from which they were 
inferred. However, it turns out that the local contributions do not. It is
possible that the missing terms are associated with contributions from the
first two (nonlocal) diagrams of Figure~\ref{diagrams} in which an $A$-type 
propagator is undifferentiated and the derivatives on the other propagator 
are contracted into one another,
\begin{equation}
\kappa a^{D-2} \times i\Delta_A(x;x') \times \partial^{\mu} \partial'_{\mu}
i\Delta(x;x') \times \kappa {a'}^{D-2} \; . \label{anomalous}
\end{equation}
In that case the contracted derivatives would produce a delta function
not recovered by the noncoincident, $D=4$ result \cite{Tsamis:1996qk},
\begin{equation}
\partial^{\mu} \partial'_{\mu} i\Delta(x;x') = -\frac{i\delta^D(x \!-\! x')}{
a^{D-2}} + O\Bigl( \frac1{\Delta x^{D-2}}\Bigr) \; . \label{anomder}
\end{equation}
It is also possible that the Feynman rules need to include contributions from
the functional measure factor. A fully dimensionally regulated calculation
would seem to be necessary to resolve this. One should also re-examine the
contribution from a loop of massless, minimally coupled scalars 
\cite{Park:2011ww} to see if it shows similar anomalous local contributions.
In the meantime, we can proceed with the nonlocal contributions because it 
turns out that the local contributions do not affect the potentials as 
strongly at late times and large distances.

\section{The Effect on the Force of Gravity}

In this section we solve the effective field equations (\ref{Einsteineqn}) to
find one loop corrections to the gravitational response to a point mass. Our
first step is to specialize the general equation (\ref{Einsteineqn}) 
appropriately for a perturbative determination of the potentials. We next
compute the source terms induced by integrating the one loop self-energy
against the classical potentials. We close the section by solving for the 
leading one loop corrections at late times and large distances. 

\subsection{Equations for the Potentials}

The linearized stress-energy for a static point mass is,
\begin{equation}
8\pi G T^{\mu\nu}_{\rm lin}(x) = 8\pi G M a \delta^3(\vec{x}) \; . \label{stress}
\end{equation}
The gravitational response to such a source is given by four scalar potentials,
\begin{equation}
\kappa h_{00} \equiv -2 \Psi \quad , \quad \kappa h_{0i} \equiv -\partial_i
\Omega \quad , \quad \kappa h_{ij} \equiv -2 \delta_{ij} \Phi - 2 \partial_i
\partial_j \chi \; . \label{potentials}
\end{equation}
We can derive an equation for $\Psi$ from the sum of the $\mu=0=\nu$ and the
spatial trace,
\begin{eqnarray}
\lefteqn{\Bigl[ \mathcal{D}^{00\rho\sigma} \!+\! \mathcal{D}^{kk\rho\sigma}\Bigr] 
\kappa h_{\rho\sigma}(x) = -2 D_B \Psi(x) = 8\pi G M a \delta^3(\vec{x}) }
\nonumber \\
& & \hspace{3cm} + \int \!\! d^4x' \Bigl\{\Bigl[\mbox{}^{00} \Sigma^{\rho\sigma}
\Bigr](x;x') \!+\! \Bigl[\mbox{}^{kk} \Sigma^{\rho\sigma}\Bigr](x;x')\Bigr\} 
\kappa h_{\rho\sigma}(x') \; , \qquad \label{Psieqn}
\end{eqnarray}
where $D_B$ is the kinetic operator of a conformally coupled scalar (\ref{DBdef}). 
The $\mu=0, \nu = i$ components give an equation for $\Omega$,
\begin{equation}
\mathcal{D}^{0i\rho\sigma} \kappa h_{\rho\sigma} = \frac{\partial_i}{2} D_B 
\Omega(x) = 0 + \int \!\! d^4x' \Bigl[\mbox{}^{0i} \Sigma^{\rho\sigma}\Bigr](x;x')
\kappa h_{\rho\sigma}(x') \; . \label{Omegaeqn}
\end{equation}
And the equation for $\chi$ and $\Psi - \Phi$ is,
\begin{eqnarray}
\lefteqn{\Bigl[ \mathcal{D}^{ij\rho\sigma} \!-\! \delta^{ij} 
\mathcal{D}^{kk\rho\sigma}\Bigr] \kappa h_{\rho\sigma} = -\partial^i \partial^j 
D_A \chi + \delta^{ij} D_A (\Psi \!-\! \Phi) = 0 } \nonumber \\
& & \hspace{2.5cm} + \int \!\! d^4x' \Bigl\{\Bigl[\mbox{}^{ij} \Sigma^{\rho\sigma}
\Bigr](x;x') \!-\! \delta^{ij} \Bigl[\mbox{}^{kk} \Sigma^{\rho\sigma}\Bigr](x;x')
\Bigr\} \kappa h_{\rho\sigma}(x') \; , \qquad \label{ChiPhieqn}
\end{eqnarray}
where $D_A$ is the kinetic operator of a massless, minimally coupled scalar
(\ref{DAdef}).

Although equations (\ref{Psieqn}-\ref{ChiPhieqn}) are correct, they cannot be
solved exactly because we only possess one loop results for the graviton 
self-energy. This means we must develop perturbative solutions,
\begin{equation}
\Psi = \Psi_0 + \kappa^2 \Psi_1 + \kappa^4 \Psi_2 + \dots \; ,
\end{equation}
and so on for the other potentials. The zeroth order solutions are,
\begin{equation}
\Psi_0(x) = \Phi_0(x) = \frac{G M}{a r} \qquad , \qquad \Omega_0(x) = \chi_0(x)
= 0 \; . \label{treepots}
\end{equation}
It is only these zeroth order potentials that appear on the right hand side
of equations (\ref{Psieqn}-\ref{ChiPhieqn}). If we use the symbol $T^i(x;x')$
to stand for just the one loop contribution to the graviton self-energy then
the one loop correction to $\Psi$ is given by,
\begin{eqnarray}
\lefteqn{ -2 D_B \kappa^2 \Psi_1(x) = \int \!\! d^4x' \Biggl\{ \Bigl[9 i T^1 \!+\!
3 i T^2 \!+\! 3 (i T^3 \!+\! i T^4) \!+\! i T^{13}\Bigr] } \nonumber \\
& & \hspace{0.5cm} + \Bigl[ 3 (i T^7 \!+\! i T^8) \!+\! i T^{12} \!+\! (i T^{16}
\!+\! i T^{17})\Bigl] {\nabla'}^2 + i T^{21} {\nabla'}^4 \Biggr\} \!\times\!
-2 \Psi_0(x') \; . \qquad \label{Psi1eqn}
\end{eqnarray}
The equations for $\Omega_1$ and $\chi_1$ are,
\begin{eqnarray}
D_B \kappa^2 \Omega_1(x) &\!\!\! = \!\!\!& \int \!\! d^4x' \Biggl\{ \Bigl[3 i T^6 
\!+\! i T^{10} \!+\! i T^{15}\Bigr] + i T^{19} {\nabla'}^2 \Biggr\} \!\times\! -2 
\Psi_0(x') \; , \qquad \label{Omega1eqn} \\
D_A \kappa^2 \chi_1(x) &\!\!\! = \!\!\!& -\!\! \int \!\! d^4x' \Biggl\{ \Bigl[3 i T^8
\!+\! i T^{12} \!+\! i T^{17}\Bigr] + i T^{21} {\nabla'}^2 \Biggr\} \!\times\! -2 
\Psi_0(x') \; . \qquad \label{Chi1eqn}
\end{eqnarray}
And the equation for the gravitational slip is,
\begin{eqnarray}
\lefteqn{ D_A \kappa^2 \Bigl[\Psi_1(x) \!-\! \Phi_1(x)\Bigr] = -\!\! \int \!\! d^4x' 
\Biggl\{ \Bigl[6 i T^1 \!+\! 2 i T^2 \!+\! 2 i T^3\Bigr] } \nonumber \\
& & \hspace{2cm} + \Bigl[ 2 i T^7 \!+\! 3 i T^8 \!+\! i T^{12} \!+\! i T^{17}\Bigr] 
{\nabla'}^2 + i T^{21} {\nabla'}^4 \Biggr\} \!\times\! -2 \Psi_0(x') \; . \qquad 
\label{Slip1eqn}
\end{eqnarray}

\subsection{Performing the Source Integrations}

From equation (\ref{Psi1eqn}) we see that $\Psi_1$ is sourced by various 
combinations of the Schwinger-Keldysh coefficient functions multiplied by zero, one 
or two powers of ${\nabla'}^2$ acing on $-2 \Psi_0(x')$. Having a factor of 
${\nabla'}^2$ simplifies the source integration enormously because,
\begin{equation}
{\nabla'}^2 \times -2 \Psi_0(x') = \frac{8 \pi G M \delta^3(\vec{x'})}{a'}
\label{nablasimp}
\end{equation}
The ${\nabla'}^4$ source comes entirely from $i T^{21}_{SK} = \frac{\kappa^2}{32 \pi^3}
\times A_3$. Table~\ref{Psinabla2} gives the combination which multiplies ${\nabla'}^2$.
\begin{table}[H]
\setlength{\tabcolsep}{8pt}
\def\arraystretch{1.5}
\centering
\begin{tabular}{|@{\hskip 1mm }c@{\hskip 1mm }||c|}
\hline
Operator & Factor\\
\hline\hline
$-12 a a' H^2 \partial_0^2 + 8 a a' (a \!-\! a') H^3 \partial_0 
- 8 a^2 {a'}^2 H^4$ & $A_1$ \\
\hline
$-18 a^2 {a'}^2 H^4 \partial_0$ & $B_1$ \\
\hline
$-12 a^3 {a'}^3 H^6$ & $C_1$ \\
\hline
$\frac{22}{3} a a' H^2 \partial_0^2 \!+\! \frac{14}{3} a a' (a \!-\! a')
H^3 \partial_0 \!-\! \frac{31}{3} a^2 {a'}^2 H^4$ & $A_2$ \\
\hline
$\frac{31}{3} a^2 {a'}^2 H^4 \partial_0 \!+\! 4 a^2 {a'}^2 (a \!-\! a')
H^5$ & $B_2$ \\
\hline
$-2 a^3 {a'}^3 H^6$ & $C_2$ \\
\hline
$-(\frac{7}{5} \partial_0^2 \!-\! \frac{107}{60} \partial^2) \partial^2
\!+\! 9 (a \!-\! a') H \partial_0 \partial^2 \!-\! \frac{37}{3} a a' H^2
\partial^2$ & $A_3$ \\
\hline
$-\ln(a) (\frac{28}{5} \partial_0^2 \!-\! \frac{107}{15} \partial^2) 
\!-\! \frac{100}{3} \ln(a) a^2 H^2$ & $\delta^4(\Delta x)$ \\
\hline
\end{tabular}
\caption{\footnotesize Contributions for $3( i T^7 \!+\! i T^8) \!+\! 
i T^{12} \!+\! (i T^{16} \!+\! i T^{17})$. Each tabulated term must be 
multiplied by $\frac{\kappa^2}{32 \pi^3}$.}
\label{Psinabla2}
\end{table}
\noindent These ${\nabla'}^2$ source integrations can be evaluated
exactly, for example, 
\begin{eqnarray}
\lefteqn{\frac{\kappa^2}{32 \pi^3} \! \int \!\! d^4x' [-8 a^2 {a'}^2 H^4] 
A_1(x;x') \times \frac{8\pi G M \delta^3(\vec{x}')}{a'} } \nonumber \\
& & \hspace{0cm} = -\frac{G M \kappa^2 H^4 a^2 \partial^2}{2 \pi^2} 
\int_{\eta_i}^{\eta - r} \!\!\!\! d\eta' a' \Bigl\{ \ln\Bigl[H^2 (\Delta 
\eta^2 \!-\! r^2)\Bigr] - 1\Bigr\} \; , \qquad \\
& & \hspace{0cm} = \frac{G M \kappa^2 H^3 a^2 \partial^2}{2 \pi^2} \Biggl\{
\ln^2\Bigl(Hr \!+\! \frac1{a}\Bigr) - \ln\Bigl(Hr \!+\! \frac1{a}\Bigl) +
\sum_{n=1}^{\infty} \frac1{n^2} \Bigl[1 - \Bigl(Hr \!+\! \frac1{a}\Bigr)^n 
\Bigr] \nonumber \\
& & \hspace{5cm} + \sum_{n=1}^{\infty} \frac1{n^2} \Bigl[ \Bigl(
\frac{Hr \!-\! \frac1{a}}{Hr \!+\! \frac1{a}}\Bigr)^n - \Bigl(Hr \!-\!
\frac1{a}\Bigr)^n \Bigr] \Biggr\} . \qquad \label{nablaexample}
\end{eqnarray}
However, all that really matters for us is the limiting form for 
$a H r \gg 1$ with $H r \ll 1$,
\begin{eqnarray}
\lefteqn{\frac{\kappa^2}{32 \pi^3} \! \int \!\! d^4x' [-8 a^2 {a'}^2 H^4] 
A_1(x;x') \times \frac{8\pi G M \delta^3(\vec{x}')}{a'} } \nonumber \\
& & \hspace{7cm} \longrightarrow \frac{2 G M \kappa^2 H^3 a^2 \ln(Hr)}{
\pi^2 r^2} \; . \qquad \label{nablaexlimit}
\end{eqnarray}

\begin{table}[H]
\setlength{\tabcolsep}{8pt}
\def\arraystretch{1.5}
\centering
\begin{tabular}{|@{\hskip 1mm }c@{\hskip 1mm }||c|}
\hline
Operator & Factor\\
\hline\hline
$-4 a a' H^2 \partial_0^4 \!+\! 28 a^2 {a'}^2 H^4 \partial_0^2 \!+\! 
24 a^2 {a'}^2 (a \!-\! a') H^5 \partial_0 \!+\! 56 a^3 {a'}^3 H^4$ & $A_1$ \\
\hline
$2 a^2 {a'}^2 H^4 \partial_0^3 \!-\! 6 a^2 {a'}^2 (a \!-\! a') H^5 \partial_0^2
\!-\! 52 a^3 {a'}^3 H^6 \partial_0$ & $B_1$ \\
\hline
$10 a^3 {a'}^3 H^6 \partial_0^2$ & $C_1$ \\
\hline
$\frac{79}{6} a a' H^2 \partial_0^4 \!+\! \frac{37}{2} a a' (a \!-\! a') H^3
\partial_0^3 \!+\! \frac{631}{2} a^2 {a'}^2 H^4 \partial_0^2$ & $A_2$ \\
$- 24 a^2 {a'}^2 (a \!-\! a') H^5 \partial_0 \!-\! 16 a^3 {a'}^3 H^6$ & \\
\hline
$-\frac{83}{2} a^2 {a'}^2 H^4 \partial_0^3 \!-\! 9 a^2 {a'}^2 (a \!-\! a')
H^5 \partial_0^2 \!+\! 34 a^3 {a'}^3 H^6 \partial_0$ & $B_2$ \\
\hline
$-9 a^3 {a'}^3 H^6 \partial_0^2$ & $C_2$ \\
\hline
$-(\frac{7}{10} \partial_0^4 \!+\! \frac14 \partial_0^2 \partial^2 \!+\!
\frac{14}{5} \partial^4) \partial^2 \!+\! \frac{29}{4} (a \!-\! a') H
(\partial_0^2 \!+\! 2 \partial^2) \partial_0 \partial^2$ & $A_3$ \\
$+ a a' H^2 (\frac{817}{12} \partial_0^2 \!+\! \frac{478}{12} \partial^2)
\partial^2 \!+\! \frac{215}{2} a a' (a \!-\! a') H^3 \partial_0 \partial^2
\!+\! \frac{315}{2} a^2 {a'}^2 H^4 \partial^2$ & \\
\hline
$-\ln(a) (\frac{14}{5} \partial_0^4 \!+\! \partial_0^2 \partial^2 \!+\!
\frac{56}{5} \partial^4) \!+\! \ln(a) a^2 H^2 (\frac{529}{3} \partial_0^2 \!+\!
\frac{334}{3} \partial^2)$ & $\delta^4(\Delta x)$ \\
$-96 \ln(a) a^3 H^3 \partial_0 \!+\! 486 \ln(a) a^4 H^4$ & \\
\hline
\end{tabular}
\caption{\footnotesize Contributions for $9 i T^1 \!+\! 3 i ^2 \!+\!
3( i T^3 \!+\! i T^4) \!+\! i T^{13}$. Each tabulated term must be 
multiplied by $\frac{\kappa^2}{32 \pi^3}$.}
\label{Psinabla0}
\end{table}

Table~\ref{Psinabla0} gives the combination of coefficient functions 
contributing to $\Psi_1(\eta,r)$ which carry no factors of ${\nabla'}^2$. 
These terms cannot be evaluated exactly, but there is no problem getting
them in the limit $a H r \gg 1$ and $H r \ll 1$. Consider the example,
\begin{eqnarray}
\lefteqn{ \frac{\kappa^2}{32 \pi^3} \! \int \!\! d^4x' [56 a^3 {a'}^3 H^4] 
A_1(x;x') \times -\frac{2 G M}{a' r'} } \nonumber \\
& & \hspace{0cm} = -\frac{7 G M \kappa^2 H^4 a^3 \partial^2}{8 \pi^3} \!
\int \!\! d^4x' \frac{{a'}^2 \theta(\Delta \eta \!-\! r')}{\Vert \vec{x}
\!+\! \vec{x}'\Vert} \Bigl\{ \ln\Bigl[ H^2 (\Delta \eta^2 \!-\! {r'}^2)
\Bigr] - 1 \Bigr\} \; , \qquad \\
& & \hspace{0cm} = -\frac{7 G M \kappa^2 H^4 a^3 \partial^2}{2 \pi^2} \!
\int_{\eta_i}^{\eta} \!\!\!\! d\eta' {a'}^2 \! \int_0^{\Delta \eta} \!\!\!\!
dr' {r'}^2 \Bigl[ \frac{ \theta(r \!-\! r')}{r} + \frac{\theta(r' \!-\! r)}{r'}
\Bigr] \nonumber \\
& & \hspace{4cm} \times \Bigl\{ \ln\Bigl[ H^2 (\Delta \eta^2 \!-\! {r'}^2)\Bigr] 
- 1 \Bigr\} \; , \qquad \\
& & \hspace{0cm} \longrightarrow \frac{7 G M \kappa^2 H^4 a^3 \partial_0^2}{
2 \pi^2 r} \! \int_{\eta_i}^{\eta} \!\!\!\! d\eta' {a'}^2 \! \int_0^{\Delta \eta} 
\!\!\!\! dr' {r'}^2 \Bigl\{ \ln\Bigl[ H^2 (\Delta \eta^2 \!-\! {r'}^2)\Bigr] 
- 1 \Bigr\} \; , \qquad \\
& & \hspace{0cm} \longrightarrow -\frac{7 G M \kappa^2 H^2 a^3 \ln^2(a)}{\pi^2 r}
\; . \qquad \label{Psi1example}
\end{eqnarray} 
When all the $\Psi_1$ source contributions are included, the leading late time
result is,
\begin{equation}
-2 D_B \kappa^2 \Psi_1 \longrightarrow -\frac{3 G M \kappa^2 H^4 a^4 [\ln^2(a) - 
\ln(Hr)]}{\pi^2 a r} + O(a^2) \; . \label{Psi1eqn2}
\end{equation} 

\begin{table}[H]
\setlength{\tabcolsep}{8pt}
\def\arraystretch{1.5}
\centering
\begin{tabular}{|@{\hskip 1mm }c@{\hskip 1mm }||c|}
\hline
Operator & Factor\\
\hline\hline
$12 a a' H^2 \partial_0^3 \!-\! 4 a a' (a \!-\! 2 a') H^3 \partial_0^2 
\!-\! 12 a^2 {a'}^2 H^4 \partial_0 \!+\! 8 a^2 {a'}^2 (a \!-\! 2a') H^5$ 
& $A_1$ \\
\hline
$10 a^2 {a'}^2 H^4 \partial_0^2 \!-\! 4 a^2 {a'}^2 (a \!+\! a') \partial_0
\!-\! 12 a^3 {a'}^3 H^6$ & $B_1$ \\
\hline
$4 a^3 {a'}^3 H^6 \partial_0$ & $C_1$ \\
\hline
$-\frac{11}{3} a a' H^2 \partial_0^3 \!-\! a a' (\frac{14}{3} a \!-\! 
\frac{74}{3} a') H^3 \partial_0^2 \!-\! 4 a^2 {a'}^2 H^4 \partial_0 \!+\!
6 a^2 {a'}^2 (a \!-\! a') H^5$ & $A_2$ \\
\hline
$\frac{53}{3} a^2 {a'}^2 H^4 \partial_0^2 \!-\! 6 a^2 {a'}^2 (a \!-\! a') H^5
\partial_0 \!-\! 12 a^3 {a'}^3 H^6$ & $B_2$ \\
\hline
$6 a^3 {a'}^3 H^6 \partial_0$ & $C_2$ \\
\hline
$(\frac75 \partial_0^2 \!-\! \frac{23}{30} \partial^2) \partial_0 \partial^2 
\!+\! \frac16 \partial^2 (83 a' \partial_0^2 \!-\! 4 a \partial^2 \!+\! 54 a' 
\partial^2) H $ & $A_3$ \\
$- \frac{91}{3} a a' H^2 \partial_0 \partial^2 \!+\! a a' (\frac{29}{3} a 
\!-\! \frac{77}{3} a') H^3 \partial^2$ & \\
\hline
$\ln(a) [(\frac{28}{5} \partial_0^2 \!-\! \frac{46}{15} \partial^2) 
\partial_0 \!+\! a H (\frac{166}{3} \partial_0^2 \!+\! \frac{100}{3} 
\partial^2) \!-\! \frac{268}{3} a^2 H^2 \partial_0 \!-\! 48 
a^3 H^3]$ & $\!\delta^4(\Delta x)\!$ \\
\hline
\end{tabular}
\caption{\footnotesize Contributions for $3 i T^6 \!+\! i T^{10} \!+\! 
i T^{15}$. Each tabulated term must be multiplied by $\frac{\kappa^2}{
32 \pi^3}$.}
\label{Omeganabla0}
\end{table}

Equation (\ref{Psi1eqn2}) shows a source term for $\Psi_1$ which grows 
like $a^3$; we ignore sources with fewer factors of $a$. Table~\ref{Omeganabla0}
gives the combinations of coefficient function which contribute to 
$\Omega_1$ and involve no factors of ${\nabla'}^2$. There is an additional 
source involving $i T^{19} \times {\nabla'}^2$. When the various source 
integrations are evaluated, and the late time form taken, the result is no 
contributions of order $a^3$,
\begin{equation}
D_B \kappa^2 \Omega_1 \longrightarrow 0 + O(a^2) \; . \label{Omega1eqn2}.
\end{equation}
Table~\ref{Chinabla0} gives the $\Omega_1$ source contributions which 
contain no factors of ${\nabla'}^2$. There is an additional contribution
involving $i T^{21} {\nabla'}^2$. When the source integrations are 
performed the result is,
\begin{equation}
D_A \kappa^2 \chi_1 = -\frac{G M \kappa^2 H^2 a^4 [5 \!-\! \ln(16)]}{8 \pi^2 a r}
+ O(a^2) \; . \label{Chi1eqn2}
\end{equation}  
\begin{table}[H]
\setlength{\tabcolsep}{8pt}
\def\arraystretch{1.5}
\centering
\begin{tabular}{|@{\hskip 1mm }c@{\hskip 1mm }||c|}
\hline
Operator & Factor\\
\hline\hline
$-8 a a' H^2 \partial_0^2 \!-\! 8 a {a'}^2 H^3 \partial_0 \!+\! 8 a^2 {a'}^2 H^4$ 
& $A_1$ \\
\hline
$-12 a^2 {a'}^2 H^4 \partial_0$ & $B_1$ \\
\hline
$-4 a^3 {a'}^3 H^6$ & $C_1$ \\
\hline
$\frac{4}{3} a a' H^2 \partial_0^2 \!-\! \frac{14}{3} a {a'}^2 H^3 \partial_0
\!+\! \frac{8}{3} a^2 {a'}^2 H^4$ & $A_2$ \\
\hline
$\frac{4}{3} a^2 {a'}^2 H^4 \partial_0 \!-\! 4 a^2 {a'}^3 H^5$ & $B_2$ \\
\hline
$-2 a^3 {a'}^3 H^6$ & $C_2$ \\
\hline
$-(\frac{7}{10} \partial_0^2 \!-\! \frac{7}{5} \partial^2) \partial^2 \!-\!
9 a' H \partial_0 \partial^2 \!-\! 5 a a' H^2 \partial^2$ & $A_3$ \\
\hline
$-\ln(a) (\frac{14}{5} \partial_0^2 \!-\! \frac{28}{5} \partial^2) \!-\! 
36 \ln(a) a H \partial_0 \!-\! 12 \ln(a) a^2 H^2$ & $\delta^4(\Delta x)$ \\
\hline
\end{tabular}
\caption{\footnotesize Contributions for $3 i T^8 \!+\! i T^{12} \!+\! 
i T^{17}$. Each tabulated term must be multiplied by $\frac{\kappa^2}{
32 \pi^3}$.}
\label{Chinabla0}
\end{table}
\noindent Tables~\ref{Slipnabla0} and \ref{Slipnabla2} give the
source combinations for the gravitational slip which contain no
factor of ${\nabla'}^2$ and one factor of it, respectively. When
the $i T^{21} \times {\nabla'}^4$ contribution is added, the
leading late time result is,
\begin{equation}
D_A \kappa^2 \Bigl[ \Psi_1 - \Phi_1\Bigr] = \frac{G M \kappa^2 H^4 a^4 
[4 \ln^2(a) \!-\! 3 \ln(Hr)]}{\pi^2 a r} + O(a^2) \; . \label{Slip1eqn2}
\end{equation}

\subsection{Solving for the Potentials}

Equations (\ref{Psi1eqn2}), (\ref{Omega1eqn2}), (\ref{Chi1eqn2}) and
(\ref{Slip1eqn2}) determine 1-loop corrections to the various potentials.
It would be straightforward to express the potentials as integrals over
the sources because we possess the exact Green's functions for $D_A$ and
$D_B$,
\begin{eqnarray}
G_A(x;x') &\!\!\! = \!\!\!& -\frac1{4\pi} \Biggl\{ \frac{\delta(\Delta \eta \!-\! 
\Delta r)}{a a' \Delta r} + H^2 \theta(\Delta \eta \!-\! \Delta r)\Biggr\} \; ,
\qquad \label{GAdef} \\
G_B(x;x') &\!\!\! = \!\!\!& -\frac1{4\pi} \frac{\delta(\Delta \eta \!-\! \Delta r)}{
a a' \Delta r} \; . \qquad \label{GBdef}
\end{eqnarray}
\begin{table}[H]
\setlength{\tabcolsep}{8pt}
\def\arraystretch{1.5}
\centering
\begin{tabular}{|@{\hskip 1mm }c@{\hskip 1mm }||c|}
\hline
Operator & Factor\\
\hline\hline
$8 a^2 {a'}^2 H^4 \partial_0^2 \!+\! 16 a^3 {a'}^2 H^5 \partial_0 
\!+\! 48 a^3 {a'}^3 H^6$ & $A_1$ \\
\hline
$-4a^3 {a'}^2 H^5 \partial_0^2 \!-\! 24 a^3 {a'}^3 H^6 \partial_0$ & $B_1$ \\
\hline
$4 a^3 {a'}^3H^6 \partial_0^2$ & $C_1$ \\
\hline
$6 a a' H^2 \partial_0^4 \!+\! \frac{37}{3} a^2 a' H^3 \partial_0^3 \!+\!
46 a^2 {a'}^2 H^4 \partial_0^2 \!-\! 16 a^3 {a'}^2 H^5 \partial_0 \!-\! 
16 a^3 {a'}^3 H^6$ & $A_2$ \\
\hline
$-24 a^2 {a'}^2 H^4 \partial_0^3 \!-\! 6 a^3 {a'}^2 H^5 \partial_0^2 \!+\! 
52 a^3 {a'}^3 H^6 \partial_0$ & $B_2$ \\
\hline
$-18 a^3 {a'}^3 H^6 \partial_0^2$ & $C_2$ \\
\hline
$(\frac{23}{60} \partial_0^2 \!-\! \frac{107}{60} \partial^2) \partial^4 \!+\!
\frac{29}{6} a H (\partial_0^2 \!+\! 2 \partial^2) \partial_0 \partial^2 \!+\!
a a' H^2 (\frac{335}{6} \partial_0^2 \!+\! \frac{65}{3} \partial^2) \partial^2$
& $A_3$ \\
$+ \frac{215}{3} a^2 a' H^3 \partial_0 \partial^2 \!+\! 56 a^2 {a'}^2 H^4
\partial^2$ & \\
\hline
$\ln(a) [(\frac{23}{15} \partial_0^2 \!-\! \frac{107}{15} \partial^2) \partial^2
\!+\! \frac{58}{3} a H (\partial_0^2 \!+\! 2 \partial^2) \partial_0 \!+\! 
a^2 H^2 (\frac{526}{3} \partial_0^2 \!+\! \frac{212}{3} \partial^2)$ & 
$\delta^4(\Delta x)$ \\
$+\frac{620}{3} a^3 H^3 \partial_0 \!+\! 176 a^4 H^4]$ & \\
\hline
\end{tabular}
\caption{\footnotesize Contributions for $6 i T^1 \!+\! 2 i T^2 \!+\! 
2 i T^3$. Each tabulated term must be multiplied by $\frac{\kappa^2}{
32 \pi^3}$.}
\label{Slipnabla0}
\end{table}
\begin{table}[H]
\setlength{\tabcolsep}{8pt}
\def\arraystretch{1.5}
\centering
\begin{tabular}{|@{\hskip 1mm }c@{\hskip 1mm }||c|}
\hline
Operator & Factor\\
\hline\hline
$-8 a a' H^2 \partial_0^2 \!-\! 8 a {a'}^2 H^3 \partial_0 \!-\! 8 a^2 
{a'}^2 H^4$ & $A_1$ \\
\hline
$-12 a^2 {a'}^2 H^4 \partial_0 \!-\! 4 a^3 {a'}^2 H^5$ & $B_1$ \\
\hline
$-8 a^3 {a'}^3 H^6$ & $C_1$ \\
\hline
$\frac{22}{3} a a' H^2 \partial_0^2 \!-\! a a' (\frac{23}{3} a \!+\! 
\frac{14}{3} a') H^3 \partial_0 \!-\! \frac{34}{3} a^2 {a'}^2 H^4$ & $A_2$ \\
\hline
$\frac{28}{3} a^2 {a'}^2 H^4 \partial_0 \!+\! 2 a^2 {a'}^2 (a \!-\! 2 a') H^5$
& $B_2$ \\
\hline
$0$ & $C_2$ \\
\hline
$-(\frac{7}{10} \partial_0^2 \!-\! \frac{107}{60} \partial^2) \partial^2 
\!+\! (\frac{29}{6} a \!-\! 9 a') H \partial_0 \partial^2 \!-\! 
\frac{19}{6} a a' H^2 \partial^2$ & $A_3$ \\
\hline
$-\ln(a) (\frac{14}{5} \partial_0^2 \!-\! \frac{107}{15} \partial^2) \!-\! 
\frac{50}3 \ln(a) a H \partial_0 \!-\! \frac{14}{3} \ln(a) a^2 H^2$ & 
$\delta^4(\Delta x)$ \\
\hline
\end{tabular}
\caption{\footnotesize Contributions for $2 i T^7 \!+\! 3 i T^8 \!+\! 
i T^{12} \!+\! i T^{17}$. Each tabulated term must be multiplied by 
$\frac{\kappa^2}{32 \pi^3}$.}
\label{Slipnabla2}
\end{table}
\noindent However, this would be overkill because the various sources are
only known for late times. It is better instead to change the temporal
variable from $\eta$ to the scale factor $a$, and then extract a factor
of $-a^4 H^2$ from the two differential operators,
\begin{eqnarray}
D_A &\!\!\! = \!\!\!& -a^4 H^2 \Bigl[ a^2 \frac{\partial^2}{\partial a^2} 
+ 4 a \frac{\partial}{\partial a} - \frac{\nabla^2}{a^2 H^2} \Bigr] \; , 
\label{newDA} \\
D_B &\!\!\! = \!\!\!& -a^4 H^2 \Bigl[ a^2 \frac{\partial^2}{\partial a^2}
+ 4 a \frac{\partial}{\partial a} + 2 - \frac{\nabla^2}{a^2 H^2} \Bigr]
\; . \label{newDB}
\end{eqnarray}
The advantage of this form is that the temporal differential operators
inside the brackets neither increase nor decrease number of scale factors,
while the effect of the spatial derivatives is sub-dominant at late times.
It is therefore trivial to invert $D_A$ and $D_B$ to the leading late time
form for the relevant sources,
\begin{eqnarray}
D_A f(a) &\!\!\! = \!\!\!& -a^4 H^2 \times \frac{[\alpha \ln^2(a) \!+\! 
\beta \ln(Hr)]}{a r} \; , \qquad \\
&\!\!\! \Longrightarrow \!\!\!& f(a) \longrightarrow -\frac{[\alpha 
\ln^2(a) \!+\! \beta \ln(Hr)]}{2 a r} \; , \qquad \label{DAinv} \\
D_B g(a) &\!\!\! = \!\!\!& -a^4 H^2 \times \frac{[\gamma \ln^2(a) \!+\!
\delta \ln(Hr)]}{a r} \; , \qquad \\
&\!\!\! \Longrightarrow \!\!\!& g(a) \longrightarrow \frac{[\frac13 \gamma
\ln^3(a) + \delta \ln(a) \ln(Hr)]}{a r} \; . \label{DBinv} \qquad 
\end{eqnarray}
Applying expression (\ref{DBinv}) to equations (\ref{Psi1eqn2}) and
(\ref{Omega1eqn2}) gives,
\begin{eqnarray}
\kappa^2 \Psi_1(\eta,r) &\!\!\! \longrightarrow \!\!\!& \frac{2 G M}{a r}
\Bigl\{ -\frac{4 G H^2 \ln^3(a)}{\pi} + \frac{12 G H^2 \ln(a) \ln(Hr)}{\pi}
\Bigr\} \; , \qquad \label{Psi1result} \\
\kappa^2 \Omega_1(\eta,r) &\!\!\! \longrightarrow \!\!\!& 0 \; . 
\label{Omega1result}
\end{eqnarray}
And the last two potentials come from using expression (\ref{DAinv}) to
invert $D_A$ in equations (\ref{Chi1eqn2}) and (\ref{Slip1eqn2}),
\begin{eqnarray}
\kappa^2 \chi_1(\eta,r) &\!\!\! \longrightarrow \!\!\!& \frac{2 G M}{a r}
\Bigl\{ \frac{[-5 \!+\! \ln(16)] G}{2 \pi} \Bigr\} \; , \qquad 
\label{Chi1result} \\
\kappa^2 (\Psi_1 \!-\! \Phi_1) &\!\!\! \longrightarrow \!\!\!& \frac{2 G M}{a r}
\Bigl\{ \frac{16 G H^2 \ln^2(a)}{\pi} - \frac{12 G H^2 \ln(H r)}{\pi} \Bigr\}
\; . \qquad \label{Slip1result}
\end{eqnarray}

\section{Epilogue}

As long as the two points do not coincide, $x^{\mu} \neq {x'}^{\mu}$,
no regularization is needed for the 1-loop graviton self-energy $-i
[\mbox{}^{\mu\nu} \Sigma^{\rho\sigma}](x;x')$. In section 2 of this
paper we exploited an old, unregulated computation of the graviton 
contribution to the graviton self-energy \cite{Tsamis:1996qk} to infer
the fully renormalized result. Our answer is expressed as a sum 
(\ref{initialrep}) of 21 coefficient functions $T^i(x;x')$, multiplied
by basis tensors listed in Table~\ref{Tbasis}. Our results for the
renormalized coefficient functions are expressed in Tables~\ref{NonABC1},
\ref{NonABC2}, \ref{NonA3}, \ref{Local} and \ref{Local4pt}, as derivative 
operators and functions of the two scale factors, acting on $\delta^4(x - x')$
and seven nonlocal functions $A_{1,2,3}(x;x')$, $B_{1,2}(x;x')$ and
$C_{1,2}(x;x')$, which are defined in expressions (\ref{ID4A}-\ref{ID4G}).

Although the nonlocal contributions obey the Ward Identity away from
coincidence, there is a local obstacle proportional to $\ln(a) \delta^4(x - x')$.
This obstacle might originate from anomalous contributions (\ref{anomalous})
to the first two diagrams of Figure~\ref{diagrams}. Such diagrams would
contribute $\ln(a) \delta^4(x - x')$ terms which we would not be able to
recognize from the unregulated, noncoincident result. It is also possible 
that we have missed some exotic, local contributions to the Feynman rules 
associated with the functional measure factor or time-ordering. More work
is required to resolve this issue, and we believe a good venue for this 
study is the much simpler contribution to $-i [\mbox{}^{\mu\nu} 
\Sigma^{\rho\sigma}](x;x')$ arising from a loop of massless, minimally 
coupled scalars \cite{Park:2011ww}. Fortunately, the missing local terms
do not make leading order contributions to the gravitational potentials.

In section 3 we applied our result to work out the gravitational response
to a static point mass (\ref{stress}) at 1-loop. Because the graviton 
self-energy was computed in a fixed gauge, we had to solve the effective 
field equations using the same gauge fixing functional \cite{Tsamis:1992xa,
Woodard:2004ut}. This resulted in there being four scalar potentials
(\ref{potentials}), instead of the usual two. Our final results for the 
leading late time forms of the four potentials were given in equations
(\ref{Psi1result}), (\ref{Omega1result}), (\ref{Chi1result}) and 
(\ref{Slip1result}). Of particular interest are the Newtonian potential
and the gravitational slip,
\begin{eqnarray}
\Psi &\!\!\! \longrightarrow \!\!\!& \frac{GM}{ar} \Biggl\{ 1 +
\frac{8 G H^2}{\pi} \Bigl[ -\ln^3(a) + 3 \ln(a) \ln(Hr)\Bigr] + \dots
\Biggr\} \; , \qquad \label{Newton} \\
\Psi - \Phi &\!\!\! \longrightarrow \!\!\!& \frac{GM}{ar} \Biggl\{ 0 +
\frac{8 G H^2}{\pi} \Bigl[4 \ln^2(a) - 3 \ln(Hr)\Bigr] + \dots \Biggr\}
\; . \qquad \label{Slip}
\end{eqnarray}

It is interesting to compare the effect of graviton contributions to the 
Newtonian potential (\ref{Newton}) with that from a loop of massless, minimally
coupled scalars \cite{Park:2015kua},
\begin{equation}
\Psi_{MMCS} \longrightarrow \frac{GM}{ar} \Biggl\{ 1 - \frac{G H^2}{10 \pi} 
\Bigl[\frac13 \ln(a) + 3 \ln(a Hr)\Bigr] + \dots \Biggr\} \; . 
\label{Newtonscalar}
\end{equation}
In both cases the 1-loop correction reduces the gravitational potential,
but gravitons induce two additional factors of $\ln(a)$. The same pattern is
evident for the gravitational slip, which gets two factors of $\ln(a)$ from
gravitons but none at all from scalars \cite{Park:2015kua}. Similarly, the
1-loop correction to the graviton mode function is enhanced by $\ln^2(a)$
\cite{Tan:2021lza}, but is not affected at all by scalars \cite{Park:2011kg}.
We therefore conclude that loops of inflationary gravitons contribute more
strongly than matter loops by two large logarithms. It is also noteworthy 
that graviton loop corrections to gravity are much strong than graviton loop
corrections to fermions \cite{Miao:2005am,Miao:2006gj,Miao:2012bj}, to
electrodynamics \cite{Leonard:2013xsa,Glavan:2013jca,Wang:2014tza,
Glavan:2015ura,Glavan:2016bvp}, and to massless, minimally coupled scalars
\cite{Kahya:2007bc,Kahya:2007cm,Glavan:2021adm}. The key difference seems to
be that graviton loop corrections to gravity can involve two graviton 
propagators whereas graviton corrections to other fields involve only one.

The appearance of very large logarithms in graviton loop corrections implies
the breakdown of perturbation at late times and large distances. It has been
difficult to devise a resummation procedure because these logarithms derive 
from two sources: the ``tail'' part of the graviton propagator and logarithmic
ultraviolet divergences of the form (\ref{renorm}) \cite{Miao:2018bol}. This
led to the speculation that resummation might be accomplished by combining a
variant of Starobinsky's stochastic formalism \cite{Starobinsky:1986fx,
Starobinsky:1994bd} with a variant of the renormalization group. This 
speculation was recently confirmed in the context of nonlinear sigma models
on a nondynamical de Sitter background \cite{Miao:2021gic}, which possess the 
same kinds of derivative interactions as quantum gravity and exhibit the same
mixture of ``tail'' and ultraviolet logarithms. The technique has been
applied to explain graviton loop corrections to the exchange potential of a
massless, minimally coupled scalar \cite{Glavan:2021adm}, and strenuous 
efforts are underway to devise similar explanations for the collection of
large graviton logarithms that have been patiently accumulated by direct 
computation over the course of two decades.

It is well known that classical modified gravity models also correct the 
force of gravity \cite{Soussa:2003re}, and can induce nonzero gravitational 
slip \cite{Nojiri:2010wj,Capozziello:2011et}. One is therefore led to wonder if 
our results (\ref{Newton}) and (\ref{Slip}) could be reproduced by some local,
metric-based model. The answer seems to be no because the only stable, local,
invariant and metric-based modification of gravity is $f(R)$ gravity 
\cite{Woodard:2006nt}. However, the modified force induced by these models on 
de Sitter background depends only on the combination $a H r$ \cite{Soussa:2003re}, 
and cannot reproduce the distinct $\ln^3(a)$ and $\ln(a) \ln(Hr)$ terms of
our result (\ref{Newton}). It should also be noted that neither the scalar nor 
the tensor amplitudes in these models experience secular growth after horizon 
crossing \cite{Brooker:2016oqa}, unlike the $\ln^2(a)$ dependence we found 
previously \cite{Tan:2021lza}.

We close by commenting on the gauge issue. On flat space background the 
graviton self-energy is known to be highly gauge dependent \cite{Capper:1973bk}.
Because the $H \rightarrow 0$ limit of our result agrees with the flat space
limit, our de Sitter graviton self-energy must inherit this gauge dependence.
The large logarithms we have found all derive from terms which carry factors
of $H^2$, and their gauge dependence is not known, although indications from
gravity plus electromagnetism suggest that there is some \cite{Glavan:2016bvp}.
A procedure has been developed for removing this gauge dependence 
\cite{Miao:2017feh}, which has been successfully applied on flat space 
background to graviton loop corrections to the massless, minimally coupled
scalar \cite{Miao:2017feh}, and to electromagnetism \cite{Katuwal:2020rkv}.
The massless, minimally coupled scalar exchange potential ha been identified
as the simplest venue for generalizing this technique to de Sitter background
\cite{Glavan:2021adm}, and it is hoped that a result will be available later
this year. Based on flat space background experience \cite{Miao:2017feh,
Katuwal:2020rkv}, we expect that the elimination of gauge dependence will not 
eliminate large graviton logarithms but might change their numerical 
coefficients.

\vspace{.5cm}

\centerline{\bf Acknowledgements}

This work was partially supported by the European Union's Seventh 
Framework Programme (FP7-REGPOT-2012-2013-1) under grant agreement 
number 316165; by the European Union's Horizon 2020 Programme
under grant agreement 669288-SM-GRAV-ERC-2014-ADG;
by NSF grant 1912484; and by the UF's Institute for Fundamental Theory.

\section{Appendix: Derivative Identities}

This Appendix summarizes the various derivative identities we employ to
convert the unregulated results of Tables~\ref{TL} and \ref{TN} to 
the renormalized Schwinger-Keldysh results of Tables~\ref{NonABC1},
\ref{NonABC2}, \ref{NonA3} and \ref{Local}.

\subsection{Extracting Derivatives}

We begin with the relations needed to write each term as derivatives 
acting on the four fundamental expressions,
\begin{equation}
\frac1{\Delta x^4} \;\; , \;\; \frac1{\Delta x^2} \;\; , \;\;
\frac{\Delta \eta}{\Delta x^2} \;\; , \;\; 
\frac{\Delta \eta^2}{\Delta x^2} \; . \label{fundamental}
\end{equation}
Terms with large inverse powers of $\Delta x^2$ all reach $\frac1{\Delta x^4}$,
\begin{eqnarray}
\frac{\Delta \eta^4}{\Delta x^{12}} = \Bigl[ \frac{\partial_0^4}{1920}
\!-\! \frac{\partial_0^2 \partial^2}{640} \!+\! \frac{\partial^4}{5120}\Bigr]
\frac1{\Delta x^4} \;\; & , & \;\; \frac{\Delta \eta^2}{\Delta x^{10}} = 
\Bigl[\frac{\partial_0^2 \partial^2}{384} \!-\! \frac{\partial^4}{1536}\Bigr]
\frac1{\Delta x^4} , \qquad \label{ID1A} \\
\frac{\Delta \eta^3}{\Delta x^{10}} = \Bigl[ \frac{\partial_0^3}{192} \!-\! 
\frac{\partial_0 \partial^2}{128} \Bigr] \frac1{\Delta x^4} \;\; & , & \;\; 
\frac{\Delta \eta}{\Delta x^{8}} = \frac{\partial_0 \partial^2}{48} \Bigl(
\frac1{\Delta x^4} \Bigr) , \qquad \label{ID1B} \\
\frac{\Delta \eta^2}{\Delta x^{8}} = \Bigl[ \frac{\partial_0^2}{24} \!-\! 
\frac{\partial^2}{48} \Bigr] \frac1{\Delta x^4} \;\; & , & \;\;
\frac{1}{\Delta x^{8}} = \frac{\partial^4}{192} \Bigl(\frac1{\Delta x^4} \Bigr) 
, \qquad \label{ID1C} \\
\frac{\Delta \eta}{\Delta x^{6}} = \frac{\partial_0}{4} \Bigl(\frac1{\Delta x^4} 
\Bigr) \;\; & , & \;\; \frac{1}{\Delta x^{6}} = \frac{\partial^2}{8} \Bigl(
\frac1{\Delta x^4} \Bigr) . \qquad \label{ID1D}
\end{eqnarray}
Terms with $\Delta \eta^4$ divided fewer than six powers of $\Delta x^2$
involve all four of the fundamental expressions (\ref{fundamental}),
\begin{eqnarray}
\frac{\Delta \eta^4}{\Delta x^{10}} & = & \frac{\partial_0^4}{384} 
\Bigl(\frac1{\Delta x^2}\Bigr) - \Bigl[ \frac{\partial_0^2}{32} \!-\!
\frac{\partial^2}{128}\Bigr] \Bigl(\frac1{\Delta x^4} \Bigr) \; , \qquad 
\label{ID1E} \\
\frac{\Delta \eta^4}{\Delta x^8} & = & \frac{\partial_0^3}{48} \Bigl( 
\frac{\Delta \eta}{\Delta x^2}\Bigr) - \frac{\partial_0^2}{8} \Bigl(
\frac1{\Delta x^2}\Bigr) + \frac18 \Bigl(\frac1{\Delta x^4} \Bigr) \; , 
\qquad \label{ID1F} \\
\frac{\Delta \eta^4}{\Delta x^6} & = & \frac{\partial_0^2}{8} \Bigl( 
\frac{\Delta \eta^2}{\Delta x^2}\Bigr) - \frac58 \partial_0 \Bigl(\frac{\Delta 
\eta}{\Delta x^2} \Bigr) + \frac38 \Bigl( \frac1{\Delta x^2}\Bigr) \; . 
\qquad \label{ID1G}
\end{eqnarray}
The last relations we require involve fewer powers of both $\Delta \eta$ and
$\Delta x^2$,
\begin{eqnarray}
\frac{\Delta \eta^3}{\Delta x^8} = \frac{\partial_0^3}{48} \Bigl(\frac1{\Delta x^2} 
\Bigr) \!-\! \frac{\partial_0}{8} \Bigl( \frac1{\Delta x^4} \Bigr) \quad & , & 
\quad \frac{\Delta \eta^3}{\Delta x^6} = \frac{\partial_0^2}{8} \Bigl( 
\frac{\Delta \eta}{\Delta x^2} \Bigr) \!-\! \frac38 \partial_0 \Bigl( 
\frac1{\Delta x^2} \Bigr) , \qquad \label{ID1H} \\
\frac{\Delta \eta^2}{\Delta x^6} = \frac{\partial_0^2}{8} \Bigl(\frac1{\Delta x^2} 
\Bigr) \!-\! \frac{1}{4} \Bigl( \frac1{\Delta x^4} \Bigr) \quad & , & \quad 
\frac{\Delta \eta^2}{\Delta x^4} = \frac{\partial_0}{2} \Bigl( 
\frac{\Delta \eta}{\Delta x^2} \Bigr) \!-\! \frac12 \Bigl( \frac1{\Delta x^2} 
\Bigr) , \qquad \label{ID1I} \\
\frac{\Delta \eta}{\Delta x^4} = \frac{\partial_0}{2} \Bigl(\frac1{\Delta x^2} 
\Bigr) \quad & , & \quad \frac{\Delta \eta^3}{\Delta x^4} = \frac{\partial_0}{2} 
\Bigl( \frac{\Delta \eta^2}{\Delta x^2} \Bigr) \!-\! \Bigl( 
\frac{\Delta \eta}{\Delta x^2} \Bigr) . \qquad \label{ID1J} 
\end{eqnarray}

\subsection{Absorbing the Factor of $\ln(H^2 \Delta x^2)$}

The next step is passing derivatives through the factor of $\ln(H^2 \Delta x^2)$
that multiplies all terms in Table~\ref{TL}. This is facilitated by the identities,
\begin{eqnarray}
\partial_0 \Bigl( \frac1{\Delta x^2}\Bigr) \!\times\! \ln(H^2 \Delta x^2) 
&\!\!\! = \!\!\!&
\partial_0 \Bigl[ \frac{\ln(H^2 \Delta x^2)}{\Delta x^2}\Bigr] + 
\frac{2 \Delta \eta}{\Delta x^4} \; , \label{ID2A} \\
\partial_0^2 \Bigl( \frac1{\Delta x^2}\Bigr) \!\times\! \ln(H^2 \Delta x^2) 
&\!\!\! = \!\!\!&
\partial_0^2 \Bigl[ \frac{\ln(H^2 \Delta x^2)}{\Delta x^2}\Bigr] + 
\frac{2}{\Delta x^4} + \frac{12 \Delta \eta^2}{\Delta x^6} \; , \label{ID2B} \\
\partial_0^3 \Bigl( \frac1{\Delta x^2}\Bigr) \!\times\! \ln(H^2 \Delta x^2) 
&\!\!\! = \!\!\!&
\partial_0^3 \Bigl[ \frac{\ln(H^2 \Delta x^2)}{\Delta x^2}\Bigr] + 
\frac{36 \Delta \eta}{\Delta x^6} + \frac{88 \Delta \eta^3}{\Delta x^8} 
\; , \label{ID2C} \\
\partial_0^4 \Bigl( \frac1{\Delta x^2}\Bigr) \!\times\! \ln(H^2 \Delta x^2) 
&\!\!\! = \!\!\!&
\partial_0^4 \Bigl[ \frac{\ln(H^2 \Delta x^2)}{\Delta x^2}\Bigr] \!+\!
\frac{36}{\Delta x^6} \!+\! \frac{528 \Delta \eta^2}{\Delta x^8} \!+\! 
\frac{800 \Delta \eta^4}{\Delta x^{10}} \; , \qquad \label{ID2D} \\
\partial_0 \Bigl( \frac{\Delta \eta}{\Delta x^2}\Bigr) \!\times\! 
\ln(H^2 \Delta x^2) &\!\!\! = \!\!\!&
\partial_0 \Bigl[ \frac{\Delta \eta \ln(H^2 \Delta x^2)}{\Delta x^2}\Bigr] + 
\frac{2 \Delta \eta^2}{\Delta x^4} \; , \label{ID2E} \\
\partial_0^2 \Bigl( \frac{\Delta \eta}{\Delta x^2}\Bigr) \!\times\! 
\ln(H^2 \Delta x^2) &\!\!\! = \!\!\!&
\partial_0^2 \Bigl[ \frac{\Delta \eta \ln(H^2 \Delta x^2)}{\Delta x^2}\Bigr] + 
\frac{6 \Delta \eta}{\Delta x^4} + \frac{12 \Delta \eta^3}{\Delta x^6} \; , 
\label{ID2F} \\
\partial_0^3 \Bigl( \frac{\Delta \eta}{\Delta x^2}\Bigr) \!\times\! 
\ln(H^2 \Delta x^2) &\!\!\! = \!\!\!&
\partial_0^3 \Bigl[ \frac{\Delta \eta \ln(H^2 \Delta x^2)}{\Delta x^2}\Bigr] \!+\!
\frac{6}{\Delta x^4} \!+\! \frac{72 \Delta \eta^2}{\Delta x^6} \!+\! 
\frac{88 \Delta \eta^4}{\Delta x^8} \; , \qquad \label{ID2G} \\
\partial_0 \Bigl( \frac{\Delta \eta^2}{\Delta x^2}\Bigr) \!\times\! 
\ln(H^2 \Delta x^2) &\!\!\! = \!\!\!&
\partial_0 \Bigl[ \frac{\Delta \eta^2 \ln(H^2 \Delta x^2)}{\Delta x^2}\Bigr] + 
\frac{2 \Delta \eta^3}{\Delta x^4} \; , \qquad \label{ID2H} \\ 
\partial_0^2 \Bigl( \frac{\Delta \eta^2}{\Delta x^2}\Bigr) \!\times\! 
\ln(H^2 \Delta x^2) &\!\!\! = \!\!\!&
\partial_0^2 \Bigl[ \frac{\Delta \eta^2 \ln(H^2 \Delta x^2)}{\Delta x^2}\Bigr] + 
\frac{10 \Delta \eta^2}{\Delta x^4} + \frac{12 \Delta \eta^4}{\Delta x^6} \; .
\label{ID2J}
\end{eqnarray}
The ``remainder'' terms, which carry no logarithms, are combined with the 
appropriate entries in Table~\ref{TN}, and then reduced to derivatives acting 
on the fundamental expressions (\ref{fundamental}) using relations 
(\ref{ID1A}-\ref{ID1J}).

\subsection{Eliminating Inverse Powers}

Reducing Table~\ref{TL} according to this scheme results in a series of 
derivatives acting on the product of a single factor of $\ln(H^2 \Delta x^2)$
times the last three terms in expression (\ref{fundamental}). The inverse
powers can be eliminated using,
\begin{eqnarray}
\frac{\ln(H^2 \Delta x^2)}{\Delta x^2} &\!\!\! \equiv \!\!\!& 2\pi i \!\times\!
A_1 = +\frac{\partial^2}{8} \Bigl[ \ln^2(H^2 \Delta x^2) \!-\! 2 
\ln(H^2 \Delta x^2) \Bigr] \; , \label{ID3A} \\
\frac{\Delta \eta \ln(H^2 \Delta x^2)}{\Delta x^2} &\!\!\! \equiv \!\!\!& 
2\pi i \!\times\! B_1 = -\frac{\partial_0}{4} \Bigl[ \ln^2(H^2 \Delta x^2) 
\Bigr] \; , \label{ID3B} \\
\frac{\Delta \eta^2 \ln(H^2 \Delta x^2)}{\Delta x^2} &\!\!\! \equiv \!\!\!& 
2\pi i \!\times\! C_1 = +\frac{\partial_0^2}{8} \Bigl[ \Delta x^2 \Bigl( 
\ln^2(H^2 \Delta x^2) \!-\! 2 \ln(H^2 \Delta x^2) \!+\! 2\Bigr) \Bigr] 
\nonumber \\
& & \hspace{6cm} + \frac14 \ln^2(H^2 \Delta x^2) \; . \qquad \label{ID3C}
\end{eqnarray}
The terms of Table~\ref{TN} produce a series of derivatives acting on the
four fundamental expressions (\ref{fundamental}). We eliminate the last three
terms using,
\begin{eqnarray}
\frac{1}{\Delta x^2} &\!\!\! \equiv \!\!\!& 2\pi i \!\times\! A_2 = 
+\frac{\partial^2}{4} \Bigl[ \ln(H^2 \Delta x^2)\Bigr] \; , \qquad 
\label{ID3D} \\
\frac{\Delta \eta}{\Delta x^2} &\!\!\! \equiv \!\!\!& 2\pi i \!\times\! B_2 = 
-\frac{\partial_0}{2} \Bigl[ \ln(H^2 \Delta x^2) \Bigr] \; , \qquad 
\label{ID3E} \\
\frac{\Delta \eta^2}{\Delta x^2} &\!\!\! \equiv \!\!\!& 2 \pi i \!\times\! 
C_2 = +\frac{\partial_0^2}{4} \Bigl[ \Delta x^2 \Bigl( \ln(H^2 \Delta x^2) 
\!-\! 1 \Bigr) \Bigr] \!+\! \frac12 \ln(H^2 \Delta x^2) \; . \qquad 
\label{ID3F}
\end{eqnarray}
The factor of $\frac1{\Delta x^4}$ is divergent. When combined with the 
appropriate counterterm it gives,
\begin{eqnarray}
\frac1{\Delta x^4} & \longrightarrow & -\frac{\partial^4}{32} \Bigl[ 
\ln^2(\mu^2 \Delta x^2) \!-\! 2 \ln(\mu^2 \Delta x^2)\Bigr] - \ln(a) \!\times\!
2\pi^2 i \delta^4(x \!-\! x') \; , \qquad \label{ID3G} \\
& \equiv & 2\pi i \!\times\! \Bigl[-\frac{\partial^2}{4} A_3 \Bigr] - \ln(a) 
\!\times\! 2\pi^2 i \delta^4(x \!-\! x') \; . \label{ID3H}
\end{eqnarray}
Note that any derivatives that act on expression (\ref{ID3H}) occur to the right
of the factor of $\ln(a)$, for example,
\begin{equation}
\partial^2 \Bigl[\frac1{\Delta x^4} \Bigr] \longrightarrow 2\pi i \!\times\! 
\Bigl[-\frac{\partial^4}{4} A_3 \Bigr] - \ln(a) \!\times\! 2\pi^2 i \partial^2 
\delta^4(x \!-\! x') \; . \label{ID3I}
\end{equation}

\subsection{Schwinger-Keldysh Reductions}

Each of the in-out logarithms in (\ref{ID3A}-\ref{ID3G}) gives rise in the
Schwinger-Keldysh formalism to real and causal expressions for $A_{1,2,3}$, $B_{1,2}$
and $C_{1,2}$,
\begin{eqnarray}
A_1 &\!\!\! \longrightarrow \!\!\!& +\frac{\partial^2}{4} 
\Biggl\{ \theta(\Delta \eta \!-\! \Delta r) \Bigl[ \ln[ H^2 (\Delta \eta^2 \!-\! 
\Delta r^2)] \!-\! 1\Bigr] \Biggr\} \; , \qquad \label{ID4A} \\
B_1 &\!\!\! \longrightarrow \!\!\!& -\frac{\partial_0}{2} 
\Biggl\{ \theta(\Delta \eta \!-\! \Delta r) \ln[ H^2 (\Delta \eta^2 \!-\! \Delta r^2)]
\Biggr\} \; , \qquad \label{ID4B} \\
C_1 &\!\!\! \longrightarrow \!\!\!& +\frac{\partial_0^2}{4} 
\Biggl\{ \theta(\Delta \eta \!-\! \Delta r) (\Delta r^2 \!-\! \Delta \eta^2) \Bigl[
\ln[ H^2 (\Delta \eta^2 \!-\! \Delta r^2)] \!-\! 1\Bigr] \Biggr\} \nonumber \\
& & \hspace{5cm} + \frac12 \theta(\Delta \eta \!-\! \Delta r)
\ln[H^2 (\Delta \eta^2 \!-\! \Delta r^2)] \; , \qquad \label{ID4C} \\
A_2 &\!\!\! \longrightarrow \!\!\!& +\frac{\partial^2}{4} 
\Biggl\{ \theta(\Delta \eta \!-\! \Delta r) \Biggr\} \; , \qquad \label{ID4D} \\
B_2 &\!\!\! \longrightarrow \!\!\!& -\frac{\partial_0}{2} 
\Biggl\{ \theta(\Delta \eta \!-\! \Delta r) \Biggr\} \; , \qquad \label{ID4E} \\
C_2 &\!\!\! \longrightarrow \!\!\!& +\frac{\partial_0^2}{4} 
\Biggl\{ \theta(\Delta \eta \!-\! \Delta r) (\Delta r^2 \!-\! \Delta \eta^2) \Biggr\} 
+ \frac12 \theta(\Delta \eta \!-\! \Delta r) \; , \qquad \label{ID4F} \\ 
A_3 &\!\!\! \longrightarrow \!\!\!& +\frac{\partial^2}{4} 
\Biggl\{ \theta(\Delta \eta \!-\! \Delta r) \Bigl[ \ln[ \mu^2 (\Delta \eta^2 \!-\! 
\Delta r^2)] \!-\! 1\Bigr] \Biggr\} \; . \qquad \label{ID4G} \\
\end{eqnarray}
Here $\Delta r \equiv \Vert \vec{x} - \vec{x}' \Vert$

\end{document}